\documentclass[a4paper,12pt]{article}
\usepackage{graphicx}
\usepackage{amsmath}
\usepackage{amssymb}
\usepackage[latin1]{inputenc}
\usepackage[english]{babel}
\usepackage{url}
\usepackage{cite}

\title{Associated charged Higgs and $W$ boson production in the MSSM
  at the CERN Large Hadron Collider} 
\author{David~Eriksson, Stefan~Hesselbach\footnote{
Present address: Institut f\"ur Theoretische Physik und Astrophysik,
Universit\"at W\"urzburg, 
D-97074 W\"urzburg, Germany}, Johan~Rathsman\\[4mm]
\normalsize
High Energy Physics, Uppsala University,
Box 535, S-75121 Uppsala, Sweden
}
\date{September 27, 2007}

\begin{document}

\maketitle

\begin{abstract}
We investigate the viability of observing charged Higgs bosons ($H^\pm$)
produced in association with $W$ bosons at the CERN Large Hadron Collider, 
using the leptonic decay $H^+ \to \tau^+ \nu_\tau$ and hadronic $W$ decay, 
within different scenarios of the Minimal Supersymmetric Standard Model (MSSM)
with both real and complex parameters.
Performing a parton level study we show how the irreducible Standard Model 
background from $W+2$ jets can be controlled by applying appropriate cuts
and find that the size of a possible signal depends on the cuts needed to
suppress QCD backgrounds and misidentifications. 
In the standard 
maximal mixing scenario of the MSSM we find a viable signal 
for large $\tan\beta$ and intermediate $H^\pm$ masses ($\sim m_t$) when using
softer cuts (\mbox{$\not \!p_\perp, p_{\perp\tau_\mathrm{jet}} >50 $ GeV})
whereas for harder cuts 
(\mbox{$\not \!p_\perp, p_{\perp\tau_\mathrm{jet}} > 100 $ GeV}) we only
find a viable signal for very large $\tan\beta$ ($\gtrsim 50$). 
We have also investigated a special class of MSSM 
scenarios with large mass-splittings among the heavy Higgs bosons 
where the cross-section can be resonantly enhanced by factors up 
to one hundred, with a strong dependence on the CP-violating phases. 
Even so we find that the signal after cuts remains small
except for small masses ($\lesssim m_t$)  
when using the softer 
cuts.
Finally, in all the scenarios we have investigated we have only found small
CP-asymmetries.

\end{abstract}

\section{Introduction}

The quest for understanding electroweak symmetry breaking and generation of
elementary particle masses is one of the driving forces behind the upcoming 
experiments at the CERN Large Hadron Collider (LHC). 
In the Standard Model (SM) the electroweak symmetry is broken by introducing
one Higgs doublet whereas in the Minimal Supersymmetric Standard Model (MSSM),
which is the main focus of this paper, two Higgs doublets are needed. 
More specifically, the MSSM is a two Higgs Doublet Model (2HDM) of type II 
where one of the doublets gives mass to all the 
up-type fermions and the other to all the down-type fermions.
After electroweak symmetry breaking the Higgs sector in the MSSM consists of
three neutral Higgs bosons and one charged.
The charged Higgs boson ($H^\pm$) is of special interest since there are no 
charged scalars in the SM and thus its discovery would constitute an 
indisputable proof of physics beyond the Standard Model.
Therefore, the hunt for charged Higgs bosons will play a central
role in the search for new physics at the LHC experiments. 

Currently the best model-independent direct limit on the charged Higgs boson 
mass is from the LEP experiments, 
$m_{H^\pm}>78.6$ GeV (95\% CL)~\cite{unknown:2001xy}
(assuming only the decays $H^+ \to c\bar s$ and $H^+ \to \tau^+\nu_\tau$).
In addition there are stronger direct limits from the Fermilab Tevatron, 
but only for very large or very small $\tan\beta$ 
($\tan\beta\gtrsim 100$ and $\tan\beta\lesssim 1$, respectively) where
$\tan\beta$ denotes the ratio of the vacuum expectation values of
the two Higgs doublets, as well as
indirect limits mainly from  B-decays although the latter are quite 
model-dependent.
We refer to the PDG~\cite{Eidelman:2004wy} for details.

The main production mode of charged Higgs bosons at hadron colliders is in
association with top quarks through the $gb \to H^-t$
and $gg \to H^-t\bar{b}$ processes~\cite{Barnett:1987jw,Bawa:1989pc,
Borzumati:1999th,Miller:1999bm} with the
former one being dominant for heavy charged Higgs bosons 
$m_{H^\pm} \gtrsim m_t$ and the latter one for light ones 
$m_{H^\pm} \lesssim m_t - m_b$. The possibilities of detecting 
charged Higgs bosons in these channels have been studied 
extensively (see for example~\cite{Gunion:1993sv,Barger:1993th,
Raychaudhuri:1995cc,Moretti:1996ra,Roy:1999xw,Moretti:1999bw,Belyaev:2002eq}
and also more specifically for the two LHC 
experiments~\cite{Biscarat,Assamagan:2002in,Kinnunen}).
In summary, the leptonic decays $H^+ \to \tau^+\nu_\tau$ seem to be most
promising for detecting light charged Higgs bosons at any value of $\tan\beta$
and heavy charged Higgs bosons for large $\tan\beta$ whereas the hadronic 
decays $H^+ \to t \bar b$ may be useful both for large and small $\tan\beta$
above threshold. However, the transition region $m_{H^\pm} \sim m_t$ has so far
been difficult to cover and the same is true for the
intermediate $\tan\beta\sim\sqrt{m_t/m_b}$ region, 
except in special supersymmetric scenarios with decays into 
charginos and 
neutralinos~\cite{Bisset:2000ud,Bisset:2003ix,Hansen:2005fg}.
The possibility of using the $gg \to H^-t\bar{b}$ process 
to improve the discovery potential in the transition region has been 
studied in~\cite{Guchait:2001pi,Assamagan:2004gv}.
Recently a new method for matching the differential cross-sections for the
two production modes has been developed~\cite{Alwall:2004xw}
resulting in a significantly improved discovery potential in the
transition region \cite{Alwall:2007}.

Given the problems with exploiting the production of charged Higgs
bosons in association with top-quarks it is also important to investigate 
other production modes. One primary example is the production 
in association with $W$ bosons~\cite{Dicus:1989vf,BarrientosBendezu:1998gd,
Moretti:1998xq,BarrientosBendezu:1999vd,BarrientosBendezu:2000tu,
Brein:2000cv,Yang:2000yt,Hollik:2001hy,Zhao:2005mu,Asakawa:2005nx}.
In addition to being a complement to the main production mode, 
especially in the transition region, this channel may also give additional 
information on the Higgs sector.
So far, $H^\pm W^\mp$ production has mainly attracted theoretical interest
with limited attention to the experimental viability, the only exception 
being~\cite{Moretti:1998xq} which came to the conclusion that in the MSSM
the signature $W^\pm H^\mp \to W^\pm tb \to W^\pm W^\mp bb$ cannot be used due to 
the irreducible background from $t \bar t$ production. On the contrary, in this
paper we will show that the situation may be drastically improved, 
at least for large $\tan\beta$, 
by instead looking at leptonic decays $H^+ \to \tau^+ \nu_\tau$ together 
with $W \to 2 \textrm{ jets}$.
Alternatively, going to a more general 2HDM one can get a very large
enhancement 
of the production cross-section compared to the MSSM
through resonant enhancement~\cite{Asakawa:2005nx}.
The associated production of charged Higgs bosons with CP-odd scalars
$A$ and subsequent leptonic $H^+ \to \tau^+ \nu_\tau$ decays have been analyzed
in \cite{Cao:2003tr}.

Different types of higher order corrections to $H^\pm W^\mp$ production 
that have been studied typically give effects of the order 10--20\%. 
This includes the supersymmetric electroweak corrections~\cite{Yang:2000yt}, 
the next-to-leading order (NLO) QCD corrections~\cite{Hollik:2001hy},
and the supersymmetric QCD corrections~\cite{Zhao:2005mu}. Of special 
importance here are the NLO QCD corrections since they show that one has to
use running masses for the Yukawa couplings, 
which in turn has a large impact on the magnitude of the cross-section as 
will be discussed later.
In addition to process specific higher order corrections there are also
higher order effects from the bottom/sbottom sector in
supersymmetric theories which are large for large $\tan\beta$ and 
therefore have to be resummed to all orders in perturbation theory
\cite{Eberl:1999he,Carena:1999py}. This results in additional terms
proportional to $\Delta m_b$ in an effective Lagrangian 
\cite{Carena:1999py} describing the couplings of charged Higgs bosons to top
and bottom quarks.
Finally  we expect that a variation of the SUSY scenario, especially of $\mu$,
will lead to a variation of the cross-section of $\mathcal{O}(10\%)$
\cite{Carena:2005ek}.
Since we will only be studying the $H^\pm W^\mp$ production at parton level,
although with appropriate smearing of momenta, we have chosen not to 
take into account any of these corrections in the production, 
except for the use of a running $b$-quark mass in the Yukawa coupling.

The study of $pp \to H^\pm W^\mp$ also offers the possibility to explore
effects of CP violation. In the MSSM with complex
parameters one gets new sources of CP violation beyond the CKM matrix which
can give rise to explicit CP violation also in the Higgs sector through
loops of supersymmetric particles~\cite{Pilaftsis:1998dd,Pilaftsis:1999qt}.
For example, such phases may give rise to differences in the 
$\mu^+\mu^- \to W^+ H^-$ and $\mu^+\mu^- \to W^- H^+$ cross-sections
which have been analyzed in a more general CP-violating 
2HDM~\cite{Akeroyd:2000zs} although not taking into account non-diagonal
Higgs propagator effects~\cite{Ellis:2004fs,Pilaftsis:1997dr}. 
Another example is the CP-odd rate asymmetry 
induced by loops of SUSY particles in the decay 
$H^+ \to t\bar{b}$~\cite{Christova:2002ke} that has been analyzed for 
the LHC in~\cite{Williams:2005gg}.

The outline of this paper is as follows: We start out by giving the
differential cross-section for the dominant production mode of 
charged Higgs bosons in association with $W$ bosons for large 
$\tan\beta$ and study the dependence on $m_{H^\pm}$ and 
$\tan\beta$ at the LHC. 
In section 3, we then specialize to the 
decays $H^+ \to \tau^+ \nu_\tau$, $W^- \to 2 \textrm{ jets}$
and compare signal and background giving special attention to the
difficult transition region where $m_{H^\pm} \sim m_t$ and the
leptonic decay mode of the charged Higgs boson is dominant.
We show how the irreducible SM background for the resulting signature
$\tau \mbox{$\not \!p_\perp$} + 2 \textrm{ jets}$
can be substantially reduced by appropriate cuts.
In section 4, we then give the results of our study in the MSSM 
both with real and complex parameters.
In the latter case we investigate the possible dependence of the signal on
some of the CP-violating phases of the MSSM. 
We also study a class of special scenarios within the MSSM
where the cross-section is resonantly enhanced.
Finally, section 5 contains our conclusions.

\section{\boldmath $H^\pm W^\mp$ production at the LHC}

In this section we give the differential cross-section for the dominant
production mode of charged Higgs bosons  in association with $W$ bosons at
large  $\tan\beta$ in a general type II 2HDM and then specialize to  the case
of the maximal mixing scenario in the MSSM giving the dependence of the total
cross-section on  $m_{H^\pm}$ and $\tan\beta$ at the LHC.

\subsection{Differential cross-section}

In a general 2HDM of type II, the dominant production mechanisms for
a charged Higgs boson in association with a $W$ boson at hadron colliders 
are $b\bar{b}$ annihilation 
\begin{equation}\label{bbprocess}
b\bar{b} \to H^\pm W^\mp
\end{equation}
at tree-level and gluon fusion 
\begin{equation}
gg \to H^\pm W^\mp
\end{equation}
at one-loop-level
\cite{Dicus:1989vf,BarrientosBendezu:1998gd}.
In this study we focus on the parameter region with intermediate 
$H^\pm$ masses ($\sim m_t$)
and large $\tan\beta$, where 
the decay $H^\pm \rightarrow \tau\nu_\tau$ has a large branching ratio and 
where the $b\bar{b}$ annihilation dominates, and thus we 
do not need to consider the contribution from gluon fusion. 
Note, that this is true also in MSSM scenarios with 
large $\tan\beta$, light
squarks and large mixing in the squark sector because the possible
strong enhancement of the
gluon fusion cross-section through squark 
loops occurs only for small $\tan\beta\lesssim 6-7$~\cite{Brein:2000cv}.

\begin{figure}
\centering
\begin{tabular}{c@{\hspace{1cm}}c}
\includegraphics{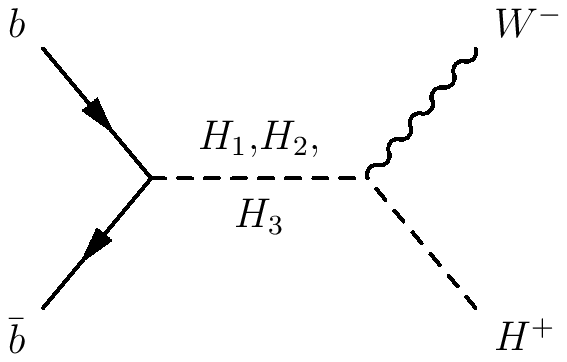} &
\includegraphics{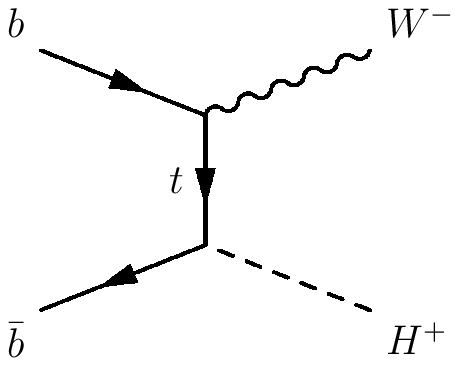} \\
\end{tabular}
\caption{Feynman diagrams for $H^\pm W^\mp$ production via
$b\bar{b}$ annihilation, $b \bar b \rightarrow H^+ W^-$. }\label{feynsignal}
\end{figure}

The Feynman diagrams for $b \bar b \rightarrow H^\pm W^\mp$ are shown
in figure~\ref{feynsignal}.
The $b\bar{b}$ annihilation proceeds either via $s$-channel exchange of 
one of the three neutral Higgs bosons in the 2HDM or $t$-channel exchange of the
top quark.
 
In the real 2HDM there are two CP-even, $h^0$ and $H^0$, and one
CP-odd, $A^0$, neutral Higgs bosons,
whereas in the general complex case all three neutral Higgs bosons mix
and form three mass eigenstates $H_i$, $i=1,\ldots 3$.
The Lagrangians relevant for the $b \bar b \rightarrow H^+ W^-$ process are,
following the conventions 
of \cite{Akeroyd:2000zs,Carena:2000yi,Lee:2003nt},
\begin{eqnarray}
\mathcal{L}_{H_i H^\pm W^\mp} & = &
  \frac{g}{2} \sum_i g_{H_i H^- W^+} (H_i \,\mathrm{i}
  \stackrel{\leftrightarrow}{\partial}_{\mu} H^-)
  W^{+,\mu} + \textrm{h.c.},\\
\mathcal{L}_{H_i \bar{b} b} & = &
  - \frac{g m_b}{2 m_W\cos\beta} \sum_i H_i \bar{b} 
    [\mathrm{Re}(g_{H_i \bar{b} b})
     - \mathrm{i} \; \mathrm{Im}(g_{H_i \bar{b} b}) \gamma_5] b, \\
\mathcal{L}_{H^\pm t b} & = &
   \frac{g}{\sqrt{2} m_W} H^+ \bar{t} \,
   [m_t \cot\beta\, \frac{1}{2}(1 - \gamma_5) 
         + m_b \tan\beta\, \frac{1}{2}(1 + \gamma_5)] \, b + \textrm{h.c.}
\end{eqnarray}
with the couplings 
\begin{equation}\label{couplings}
\begin{aligned}
g_{H_i H^- W^+}&= g_{H_i H^+ W^-}^* = 
  O_{2i}\cos\beta-O_{1i}\sin\beta+ \mathrm{i} \, O_{3i} \,,\\
g_{H_i \bar{b} b}&=O_{1i}+ \mathrm{i} \, O_{3i}\sin\beta \,,
\end{aligned}
\end{equation}
where $O_{ji}$ is the Higgs mixing matrix. 
In the real 2HDM, $H_i=\{h^0,H^0,A^0\}$ and the mixing 
matrix has the simple form
\begin{equation}
O_{ji}=
\begin{pmatrix}
-\sin\alpha & \cos\alpha & 0 \\
\cos\alpha & \sin\alpha & 0 \\
0 & 0 & 1 \\
\end{pmatrix}
\end{equation}
which gives purely real couplings for $h^0$, $H^0$ and imaginary ones for $A^0$.

In the general CP-violating 2HDM all entries in the mixing matrix 
can be non-zero. The
resulting mixing between CP-even and CP-odd Higgs states leads to  
complex couplings in general,
which together with the propagators gives differences between 
the cross-sections for $b\bar b \to H^+ W^-$ and $b\bar b \to H^- W^+$ 
in analogy to the $\mu^+\mu^- \to W^\pm H^\mp$ processes analyzed 
in~\cite{Akeroyd:2000zs}. However, it has to be kept in mind that the
formalism we are using is simplified and only valid if at most one of 
the neutral Higgs propagators can be resonant. Otherwise one has to use 
the complete description~\cite{Ellis:2004fs,Pilaftsis:1997dr}, which also takes 
into account 
non-diagonal propagator effects that arise from the mixing of two different 
Higgs bosons through higher order loops.
Using the simplified formalism, the differential cross-sections for the two
processes are~\cite{BarrientosBendezu:1998gd,Akeroyd:2000zs}: 
\begin{multline}
\frac{d\sigma}{dt}(b \bar b \rightarrow H^+ W^-)=
\frac{G_F^2}{ 24\pi s}\\
\left\{{m_b^2 \lambda(s,m_W^2,m_{H^\pm}^2)\over 2\cos^2\beta}
\sum_{i,j}g_{H_i H^- W^+}g_{H_j H^- W^+}^*S_{H_i}S_{H_j}^*
\mathrm{Re}[g_{H_i \bar b b}g_{H_j \bar b b}^*]\right.\\
+{1\over (t-m_t^2)^2}\left[m_t^4 \cot^2\beta (2m_W^2+p_\perp^2)
+m_b^2 \tan^2\beta(2m_W^2p_\perp^2+t^2)\right]\\
\left.+{m_b^2 \tan\beta\over (t-m_t^2)\cos\beta}\left[m_W^2 m_{H^\pm}^2-s p_\perp^2-t^2\right]
\sum_{i}\mathrm{Re}\left[g_{H_i H^- W^+}g_{H_i \bar b
    b}S_{H_i}\right]\right\} ,
\label{eq:bbhpwm}
\end{multline}
\begin{multline}
\frac{d\sigma}{dt}(b \bar b \rightarrow H^- W^+)=
{G_F^2\over 24\pi s}\\
\left\{{m_b^2 \lambda(s,m_W^2,m_{H^\pm}^2)\over 2\cos^2\beta}
\sum_{i,j}g_{H_i H^- W^+}^*g_{H_j H^- W^+}S_{H_i}S_{H_j}^*
\mathrm{Re}[g_{H_i \bar b b}^*g_{H_j \bar b b}]\right.\\
+{1\over (t-m_t^2)^2}\left[m_t^4 \cot^2\beta (2m_W^2+p_\perp^2)
+m_b^2 \tan^2\beta(2m_W^2p_\perp^2+t^2)\right]\\
\left.+{m_b^2 \tan\beta\over (t-m_t^2)\cos\beta}\left[m_W^2 m_{H^\pm}^2-s p_\perp^2-t^2\right]
\sum_{i}\mathrm{Re}\left[g_{H_i H^- W^+}^*g_{H_i \bar b
    b}^*S_{H_i}\right]\right\} ,
\label{eq:bbhmwp}
\end{multline}
where $s$ and $t$ are the ordinary Mandelstam variables of the hard process and
\begin{equation}
\lambda(x,y,z)=x^2+y^2+z^2-2(xy+yz+zx) \, ,
\end{equation}
\begin{equation}
p_\perp^2 = \frac{\lambda(s,m_W^2,m_{H^\pm}^2) \sin^2\theta}{4s}  \, ,
\end{equation}
with $\theta$ being the polar angle in the $2\to2$ cms and
\begin{equation}
S_{H_i}={1\over s - m_{H_i}^2 + i m_{H_i} \Gamma_{H_i}} .
\label{eq:SHi}
\end{equation}

In the remainder of this paper we focus on $H^\pm W^\mp$ production in
the MSSM, with real and complex parameters, 
which is a special case of a type II 2HDM.
The Higgs masses, widths, branching ratios
and the mixing matrix of the neutral Higgs bosons in the 
MSSM can be calculated with programs such as {\sc FeynHiggs}
\cite{Heinemeyer:2001qd,Hahn:2005cu,feynhiggs,Frank:2006yh} and
{\sc CPsuperH} \cite{Lee:2003nt}, which include higher order corrections up
to leading two-loop contributions, especially the $\Delta m_b$
corrections.

As already alluded to, $H^\pm W^\mp$ production could
potentially be used to determine the amount of CP violation in the MSSM. In
order to isolate the possible effects of CP violation, a CP-odd rate asymmetry
\begin{equation}\label{eq:asymmetry}
A_\mathrm{CP} = \frac{\sigma(pp \rightarrow H^+ W^-) - 
  \sigma(pp \rightarrow H^- W^+)}
  {\sigma(pp \rightarrow H^+ W^-) + 
  \sigma(pp \rightarrow H^- W^+)}
\end{equation}
can be defined,
where only interference 
terms contribute to the numerator
\begin{multline}
\sigma(pp \rightarrow H^+ W^-) - \sigma(pp \rightarrow H^- W^+)
  \propto 
\frac{d\sigma}{dt}(b \bar b \rightarrow H^+ W^-) - \frac{d\sigma}{dt}(b \bar b \rightarrow H^- W^+)
 = \\ =
{G_F^2\over 12\pi s} 
 \Bigg\{ {m_b^2 \lambda(s,m_W^2,m_{H^\pm}^2)\over \cos^2\beta}
 \sum_{i > j} 
  \mathrm{Im}(g_{H_i H^- W^+}g_{H_j H^- W^+}^*)
  \mathrm{Im}(S_{H_i}S_{H_j}^*)
  \mathrm{Re}(g_{H_i \bar b b} g_{H_j \bar b b}^*) \\
 + {m_b^2 \tan\beta\over (t-m_t^2)\cos\beta}
    \left[m_W^2 m_{H^\pm}^2-s p_\perp^2-t^2\right]
     \sum_{i}
    \mathrm{Im} ( g_{H_i H^- W^+}g_{H_i \bar b b} ) 
    \mathrm{Im} ( S_{H_i} )
\Bigg\}.
\label{eq:asydenom}
\end{multline}
From this expression it is clear that the conditions for obtaining large CP
asymmetries are quite subtle. Starting with the first term, which comes from the
interference of two different $s$-channel amplitudes, we see that in
this case large asymmetries can be obtained only if the 
phases in the propagators $S_{H_i}$ make $ \mathrm{Im}(S_{H_i}S_{H_j}^*) $
large and at the same time the Higgs mixing matrix  $O_{ij}$ leads to large
factors from the couplings 
$\mathrm{Im}(g_{H_i H^- W^+}g_{H_j H^- W^+}^*) $ and
$\mathrm{Re}(g_{H_i \bar b b} g_{H_j \bar b b}^*) $. Typically this means that
the maximal asymmetry with one resonant propagator is of order 
$\Gamma_{H_i}/m_{H_i}$, modulo the coupling factors. 
In the more general case with several resonant propagators the formalism we 
are using is not sufficient as already noted. 
Looking at the second term in
Eq.~(\ref{eq:asydenom}), which originates in the interference of one 
$s$-channel
amplitude with the $t$-channel one, this can give an asymmetry of the order
$\Gamma_{H_i}/m_{H_i}$ if the $s$-channel amplitude is resonant, again modulo
the coupling factor. However, in this case the coupling factor will make the
asymmetry suppressed with $1/\tan\beta$ for large $\tan\beta$ 
compared to the asymmetry arising from the interference of two $s$-channel
amplitudes.

\subsection{Cross-section calculation}

We have implemented~\cite{DEriksson} the two processes $b\bar b \rightarrow H^+ W^-$ 
and $b\bar b \rightarrow H^- W^+$ as separate external processes to
{\sc Pythia}~\cite{Sjostrand:2000wi,Sjostrand:2003wg}  
in order to be able to analyze the rate asymmetry $A_\mathrm{CP}$.
In principle, the implementation in {\sc Pythia} makes a generation of
the complete 
final state possible, but for this first study we have chosen to stay
on leading 
order parton level.

We use running $b$ and $t$ masses in the Yukawa
couplings and a running electroweak coupling $\alpha_{EW}$, all
evaluated at the renormalisation scale $\mu_R= m_{H^\pm}+m_{W}$
using the leading order formulas implemented in {\sc Pythia}. 
Numerically, for $ m_{H^\pm}=175 (400)$ GeV the masses are 
$m_b=2.7 (2.6)$ and $m_t=160 (153)$ GeV.
This leads to a reduction of the cross-section compared to 
the naive tree-level without running to about a third, which is 
in better agreement with the NLO calculation \cite{Hollik:2001hy}.
As factorization scale we use the average mass of the $H^\pm$ and the
$W$, $\mu_F=\frac{1}{2}(m_{H^\pm}+m_{W})$, 
again for better agreement with NLO calculations. 
All other parameters are left at their default values in {\sc Pythia},
for example, we use the CTEQ 5L~\cite{Lai:1999wy} parton densities.
We note that the main uncertainty in the $b$-quark parton density is not 
so much a question of which set that is used, but rather the precise
value of the factorisation scale. Varying the scale with a factor 2 up or down
the cross-section changes with $\pm20$\%. This should be compared with the 
about $\pm5$\% uncertainty in the $b$ density given for example by
CTEQ65E~\cite{Tung:2006tb} for relevant values of $x$ and $Q^2$.\footnote{For 
this estimate we 
have used the online tool for pdf plotting and calculation at 
\url{http://durpdg.dur.ac.uk/hepdata/pdf3.html}.}
In addition the widths of the $H^\pm$ and $W$ bosons are also
included in the same way as in standard {\sc Pythia}. 
In other words the $m_{H^\pm}$ and $m_{W}$ masses vary according to
Breit-Wigner distributions with varying widths meaning that 
for each mass the decay widths are recalculated based on the actually open
decay channels.

For the calculation of the MSSM scenario and the corresponding Higgs
masses, the Higgs mixing matrix $O_{ji}$ and the branching ratios of $H^\pm$ 
we use {\sc FeynHiggs} 2.2.10 \cite{feynhiggs}.
From $O_{ji}$ we then calculate the couplings $g_{H_i H^- W^+}$ and 
$g_{H_i \bar{b} b}$ according to Eq.~(\ref{couplings}).

\begin{figure}
\centering
\begin{tabular}{cc}
\includegraphics[width=7.5cm]{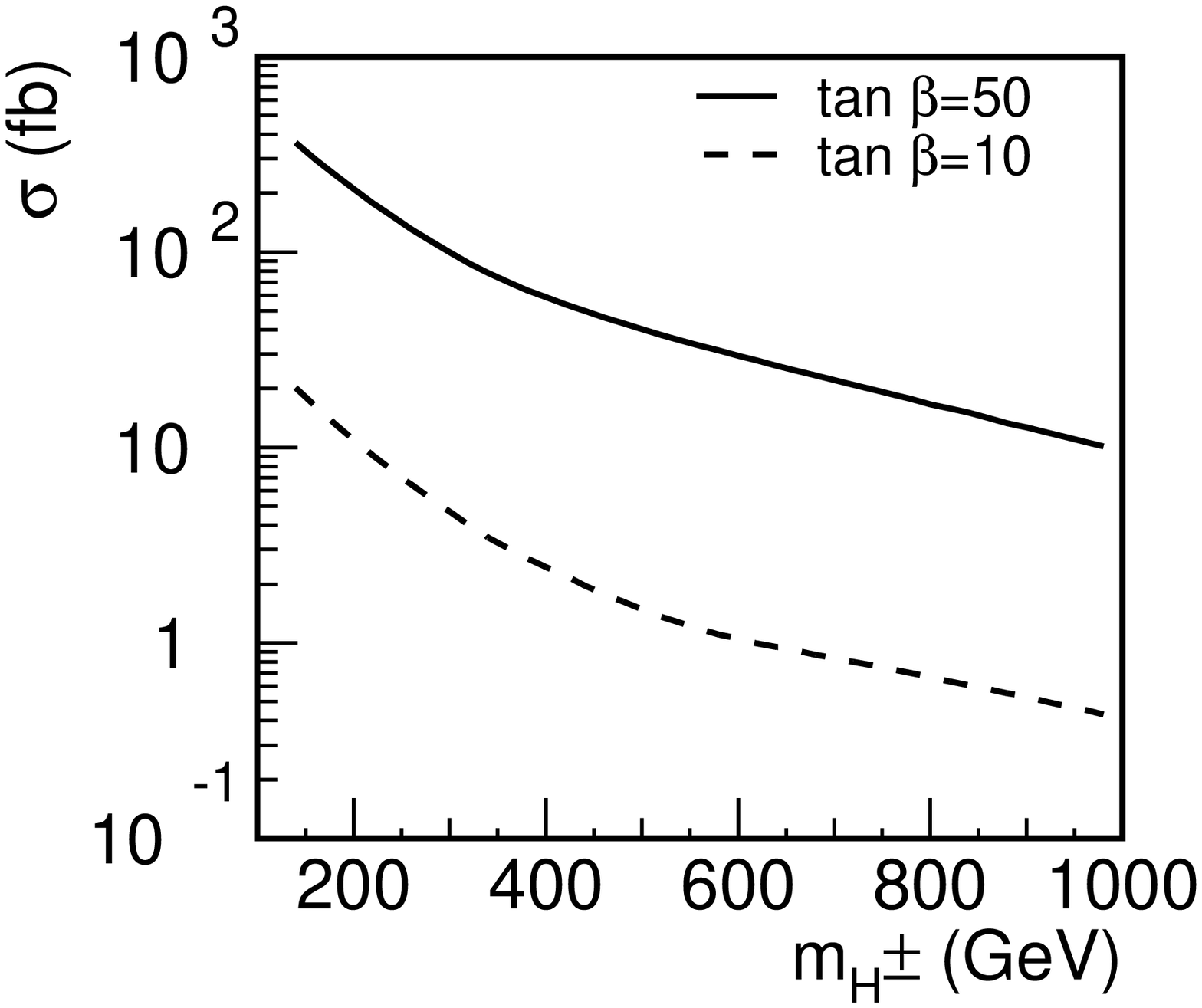} &
\includegraphics[width=7.5cm]{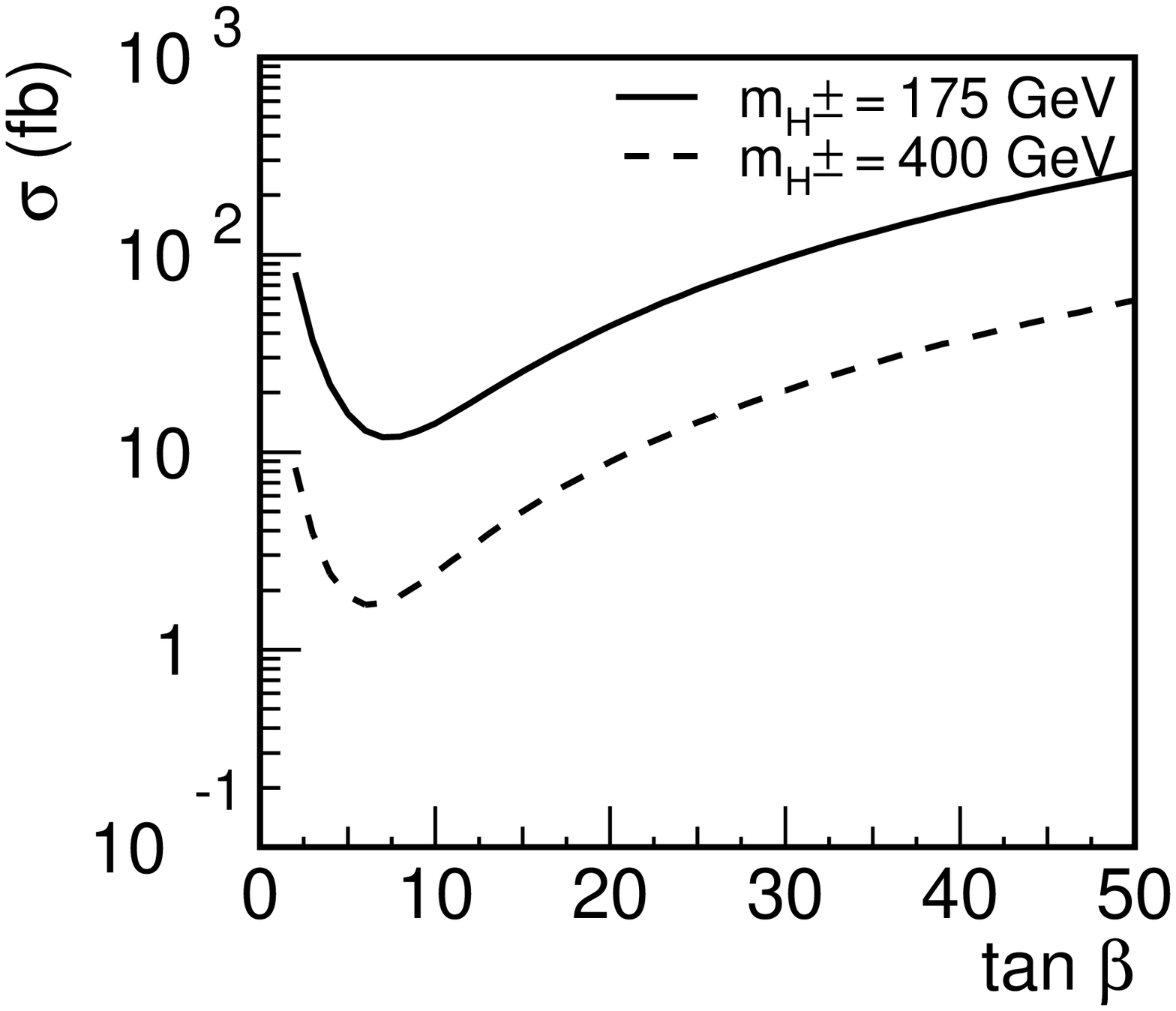} \\
\includegraphics[width=7.5cm]{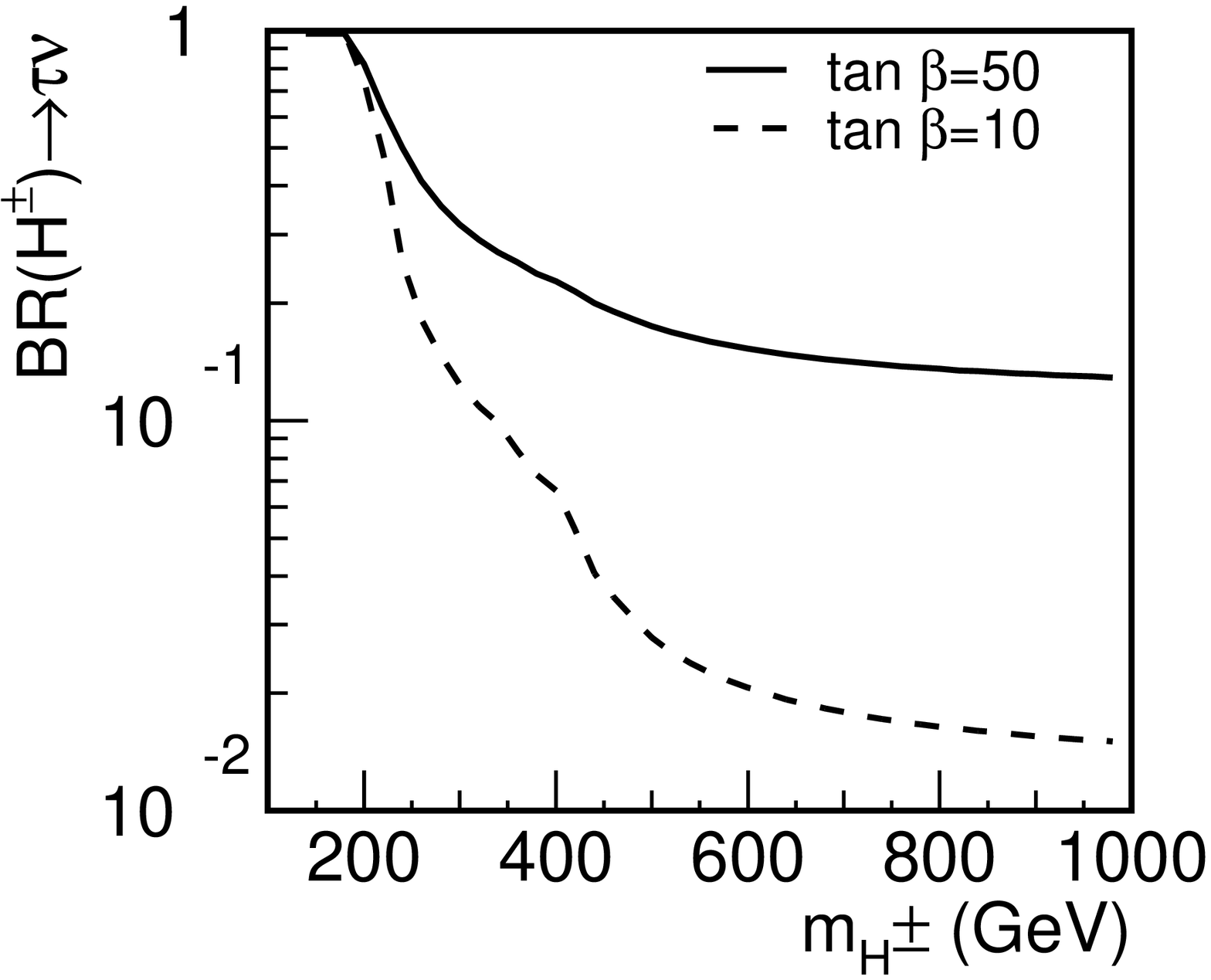} &
\includegraphics[width=7.5cm]{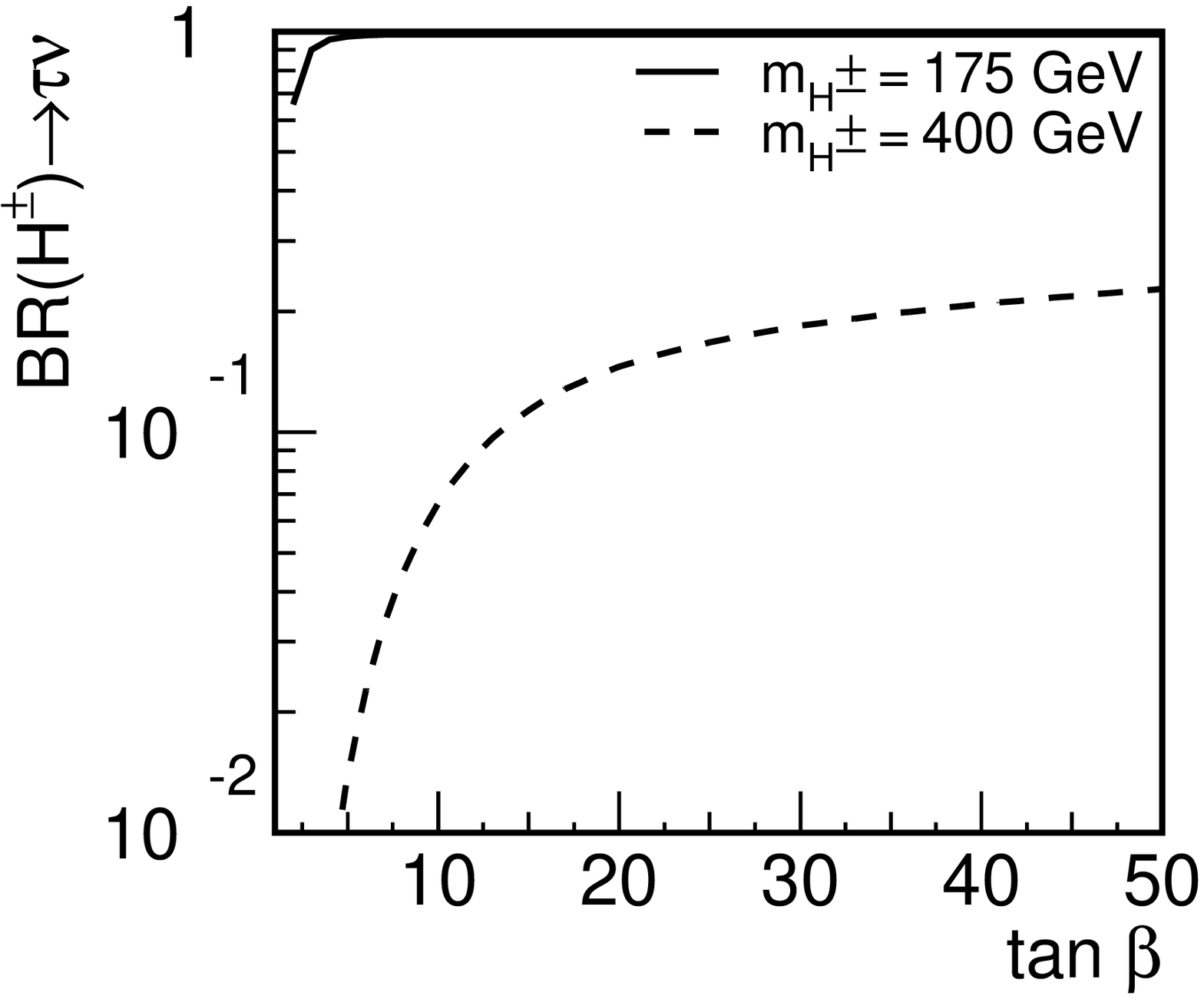} \\
\end{tabular}
\caption{The left plots show the ($b\bar b$ contribution to the) cross-section 
$\sigma(pp \to H^\pm W^\mp)$ at the LHC
and the branching ratio $BR(H^\pm \to \tau\nu_\tau)$ as a
function of $m_{H^\pm}$ for $\tan\beta=50$ (solid line) and
$\tan\beta=10$ (dashed line).
The right plots show the cross-section and branching ratio as a
function of $\tan\beta$ for $m_{H^\pm}=175$ GeV (solid line) and 
$m_{H^\pm}=400$~GeV (dashed line).}\label{fig:sigma_mhc_tanbeta} 
\end{figure}

Figure~\ref{fig:sigma_mhc_tanbeta} shows the resulting 
 total cross-section $\sigma(pp \to H^\pm W^\mp)$
as well as the $H^\pm$ branching ratio into $\tau\nu_\tau$ as functions of 
the $H^\pm$ mass and $\tan\beta$ at the LHC. 
Here, and in the following unless otherwise noted, we have used a maximal
mixing scenario, defined in table~\ref{maxmixpar}\footnote{%
Here $\tan\beta \equiv v_2/v_1$ denotes the ratio of the vacuum
expectation values of the Higgs fields and $\mu$ is the Higgs/higgsino
mass parameter. $M_\mathrm{SUSY}$ defines a common value of all squark
and slepton soft SUSY breaking
mass parameters $M_Q^g$, $M_U^g$, $M_D^g$, $M_L^g$,
$M_E^g$, where $g=1,2,3$ is the generation index.
$X_t= A_t - \mu^*\cot\beta$ and $X_b = A_b - \mu^* \tan\beta$ 
describe the mixing in the third generation squark sector
with the trilinear scalar couplings  $A_t$ and $A_b$.
In the gaugino sector $M_2$ denotes the SU(2) soft SUSY breaking mass
parameter, $m_{\tilde{g}}$ is the mass of the gluinos and the U(1)
soft SUSY breaking mass parameter $M_1$ is fixed by the GUT relation
$M_1/M_2 = 5/3 \tan^2\theta_W$.}.
At the same time, we have found that the cross-section depends 
very little on the mixing in the third generation squark sector.
For example, for a mass of $m_{H^\pm}=175$~GeV and $\tan\beta=50$ the
cross-section decreases from 260~fb to 257~fb in the no mixing scenario of
table~\ref{maxmixpar}.

\begin{table}
\centering
\begin{tabular}{lcccc|cccc} \hline \hline
& \multicolumn{7}{c}{MSSM Parameters.
  All masses in GeV.}\\
Scenario & 
$A_t$ & 
$A_b$ & 
$M_2$ &  
$m_{\tilde g}$   &  
$m_{\tilde t_1}$  &
$m_{\tilde t_2}$  &
$m_{\tilde b_1}$  &
$m_{\tilde b_2}$  
\\ \hline
Maximal mixing ($m_h^\mathrm{max}$) &
2000 & 2000 & 200 & 800 & 820  & 1177  & 996  & 1012 \\
Less mixing  &
1000 & 1000 & 200 & 800 & 922  & 1099  & 993  & 1012  \\
No mixing  &
0 & 0 & 200 & 800  & 1014  & 1015  & 991  & 1014 \\
\hline \hline
\end{tabular}
\caption{MSSM parameters for the maximal mixing scenario,
  $m_h^\mathrm{max}$, the less mixing scenario and the no mixing
  scenario in addition to $\tan\beta=50$,
$\mu=200$ GeV , and 
$M_\mathrm{SUSY}=1000$ GeV, as well as the 
resulting 3rd generation squark masses.}\label{maxmixpar}
\end{table}

\begin{figure}
\centering
\includegraphics[width=7.5cm]{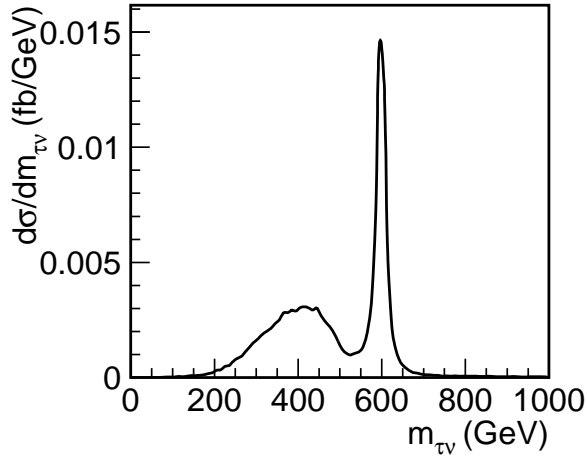}
\caption{Distribution of the invariant mass of the decay products
(in our case $\tau
\nu_\tau$) from the $H^\pm$ showing the off-shell enhancement of the
cross-section for $m_{H^\pm}=600$ GeV, resulting in
$m_{H_{1,2,3}} = 136, 594, 594$ GeV. 
In this case the large width 
$\Gamma_{H^\pm}(m_{H^\pm})=20.3$ GeV means that the mass can fluctuate down to 
$m_{\tau\nu_\tau}+m_W<m_{H_i}$ which results in the broad peak seen to the left.}
\label{fig:offshell}
\end{figure}

Apart from the decreased cross-section due to the use of a running mass instead
of the pole mass, our results also differ from, for example, the ones by 
\cite{BarrientosBendezu:1998gd} at large Higgs masses due to off-shell 
resonant enhancement.
When the width of the $H^\pm$ is large, the mass of the decay products 
(in our case $\tau \nu_\tau$)  can fluctuate down such that
$m_{\tau\nu_\tau}+m_W<m_{H_i}$ giving rise to 
$s$-channel resonant production. 
This can be seen from the big broad peak
in the $m_{\tau\nu_\tau}$ invariant mass spectrum shown in 
figure~\ref{fig:offshell}, 
which has been obtained for the following masses:
$m_{H^\pm}=600$ GeV, $m_{H_1}=136$ GeV, $m_{H_2}=594$ GeV, and $m_{H_3}=594$ GeV.
In turn this leads to an enhancement of the cross-section for large Higgs
masses. The same effect can also be seen in the study by~\cite{Moretti:1998xq}.
However, 
it is not clear to what extent this increase in the cross-section leads to a
larger signal in the end since the much wider peak
could potentially be harder to see. This is especially true for the leptonic
decays we are considering since here we can at best reconstruct the sum 
of the transverse momenta carried by neutrinos.

\section{Signal selection}

In this section we explain the final signature analyzed in this paper,
discuss background processes and define the cuts necessary to enhance
the signal-to-background ratio.

The signal selection is based on the intermediate mass $m_{H^\pm}=175$ GeV
for $\tan\beta=50$ in the maximal mixing scenario given in 
table~\ref{maxmixpar}. Results based on this selection for other values 
of the mass and $\tan\beta$ as well as in other scenarios will be given in the
next section.

\subsection{Signature}

As already mentioned, in this study we focus on 
associated $H^\pm W^\mp$ production with subsequent leptonic
decays of the charged Higgs, $H^+ \to \tau^+ \nu_\tau$ and hadronic decays
of the $W$ boson, $W^- \to \bar{q} q'$, $q=u,c$, $q'=d,s$.
The  decays $W^+ \to \tau^+ \nu_\tau$ and  
$H^- \to \bar{c} s,  \bar{c} b$ have not been
included since the branching ratios 
for the latter are negligible compared to $H^+ \to \tau^+ \nu_\tau$ 
for large $\tan\beta$.
For simplicity we only consider hadronic decays of the $\tau$-lepton, 
$\tau \to \nu_\tau + \mathrm{hadrons}$. 
The resulting signature thus consists of two light jets ($j$),
one hadronic $\tau$ jet ($\tau_\mathrm{jet}$) and missing transverse momentum
(\mbox{$\not \!p_\perp$}) carried away by the two neutrinos (one from the
charged Higgs boson decay and one from the $\tau$-lepton decay):
$$2 j + \tau_\mathrm{jet} + \mbox{$\not \!p_\perp$} \, .$$

Due to the two neutrinos escaping detection it is of course not possible to
reconstruct the invariant mass of the charged Higgs boson in this decay mode.
However,
from \mbox{$\not \!p_\perp$} and the transverse momentum of the $\tau$
jet, $p_{\perp\tau_\mathrm{jet}}$, the transverse mass\footnote{Strictly
speaking this is not the transverse mass since there are two neutrinos in the
decay chain of the charged Higgs boson we are considering. Even so the
characteristics of this mass is very similar to that of the
true transverse mass.} 
\begin{equation}
m_\perp=
\sqrt{2p_{\perp\tau_\mathrm{jet}} \mbox{$\not \!p_\perp$} [1-\cos(\Delta\phi)]}
\end{equation}
can be calculated,
where $\Delta\phi$ is the azimuthal angle between $p_{\perp\tau_\mathrm{jet}}$ 
and \mbox{$\not \!p_\perp$}.
If there is a detectable charged Higgs boson it will show up as a peak in
this distribution with the upper edge of the peak given by the 
mass of the charged Higgs boson.

\subsection{Background}

The dominant irreducible SM background for our signature
$2 j + \tau_\mathrm{jet} + \mbox{$\not \!p_\perp$}$ arises from $W$ + 2 jets
production which we have simulated with help of the package 
ALPGEN \cite{Mangano:2002ea}. The program  calculates the exact matrix 
elements on tree-level for the $2 j + \tau + \nu_\tau$ final state. 
In this way it includes not only $W + 2 \textrm{ jets}$ production 
as well as $W$ pair production,
but also contributions where the $\tau$ lepton and the
neutrino do not arise from the decay of a (virtual) $W$ boson.
The obtained cross-sections and distributions of momenta,
invariant masses etc.\ have been cross-checked with Madgraph
\cite{Murayama:1992gi,Stelzer:1994ta,Maltoni:2002qb}.

As a  further precaution 
we have also simulated another irreducible background,
namely $WZ + 2~\mathrm{jets}$ production with $Z \to \nu\nu$ using ALPGEN, 
which typically has a larger missing transverse momentum. However, 
we have found that this process
contributes less than 3\% to the overall background
after appropriate cuts (table~\ref{tbcuts} below) have been applied.
So even though the tails of the $p_\perp$ and $m_\perp$ distributions 
are slightly harder compared to the $W + 2 \textrm{ jets}$ background, it is
still safe to neglect the $WZ + 2~\mathrm{jets}$ background for this study.

\begin{figure}
\centering
\includegraphics[width=7.5cm]{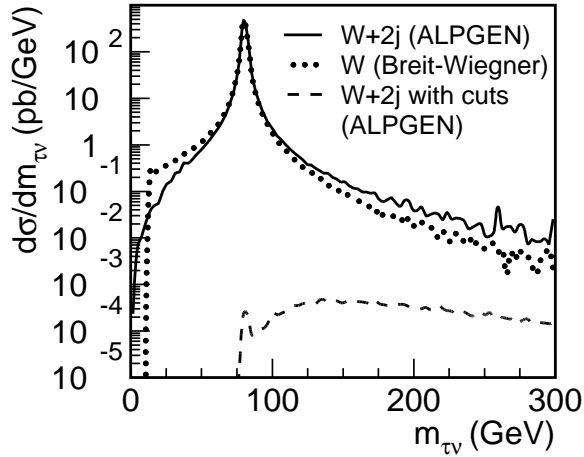}
\caption{Invariant mass $m_{\tau\nu_\tau}$
of the $\tau$ and $\nu_\tau$ system generated with ALPGEN (solid line) compared
to the
$W$ mass from {\sc Pythia} in $W + \textrm{jet}$ production (dotted line),
which has a normal Breit-Wigner distribution with varying width. 
The distributions have been normalised to the same height. Also shown is
the $m_{\tau\nu_\tau}$ distribution generated with ALPGEN after applying 
the cuts defined in table \ref{tbcuts} (dashed line).}\label{compare_mw}
\end{figure}

The importance of using the complete matrix element for the signature we are
considering, and not just the $Wjj$ approximation,
can be seen in figure~\ref{compare_mw}.
The figure shows the invariant mass $m_{\tau\nu_\tau}$
of the $\tau$ and $\nu_\tau$ system generated with ALPGEN
compared to the $W$ mass from {\sc Pythia} in $W + \textrm{jet}$ production,
which has a normal Breit-Wigner distribution with varying width.
Of special interest here is the tail for large invariant masses 
$m_{\tau\nu_\tau}$ and we note that 
there is a factor of 2 difference between the two approaches in this tail.
To illustrate the invariant mass range of relevance in our case 
we also show $m_{\tau\nu_\tau}$ generated with ALPGEN
after all event selection cuts (see table~\ref{tbcuts})
have been applied.
This also shows that the SM background can be considerably reduced by
appropriate cuts, which will be discussed in the next subsection.

\subsection{Cuts}

Our study is performed at parton level, without any parton showering or
hadronisation. Instead the momenta of the jets are smeared as a first 
approximation to take these, as well as detector effects, into account. 
The only exception is that for both signal and background
{\sc Tauola}~\cite{Jadach:1990mz,Golonka:2003xt} has 
been used to perform the decay of the $\tau$ lepton into a hadronic jet plus a
neutrino. The use of {\sc Tauola} also makes it possible to take into
account the differences in the decay characteristics of the $\tau$-lepton 
depending on whether
it comes from a charged Higgs bosons with spin 0 or a W-boson with spin 1. 
This difference can, at least in principle, also be used to discriminate 
against 
backgrounds~\cite{Raychaudhuri:1995cc,Roy:1999xw,Roy:1991sf,Guchait:2006jp}
as discussed below.

We distinguish between the two jets by calling them hard (with momentum
$p_{hj}$) and soft ($p_{sj}$) according to the larger and smaller value
of their transverse momentum $p_\perp$, respectively. 
We have done a Gaussian smearing of the measurable momenta
$p_{\tau_\mathrm{jet}}$, $p_{hj}$ and $p_{sj}$ and then calculated
\mbox{$\!\not\!p_\perp$}. 
The smearing is done preserving the direction of the momenta and the
width of the smearing is defined as
$\sigma[\mathrm{GeV}] = \sqrt{{a^2 \over p_\perp[\mathrm{GeV}]} + b^2}$
with $a=0.60$ and $b=0.02$.

\begin{figure}
\centering
\begin{tabular}{cc}
\includegraphics[width=7.5cm]{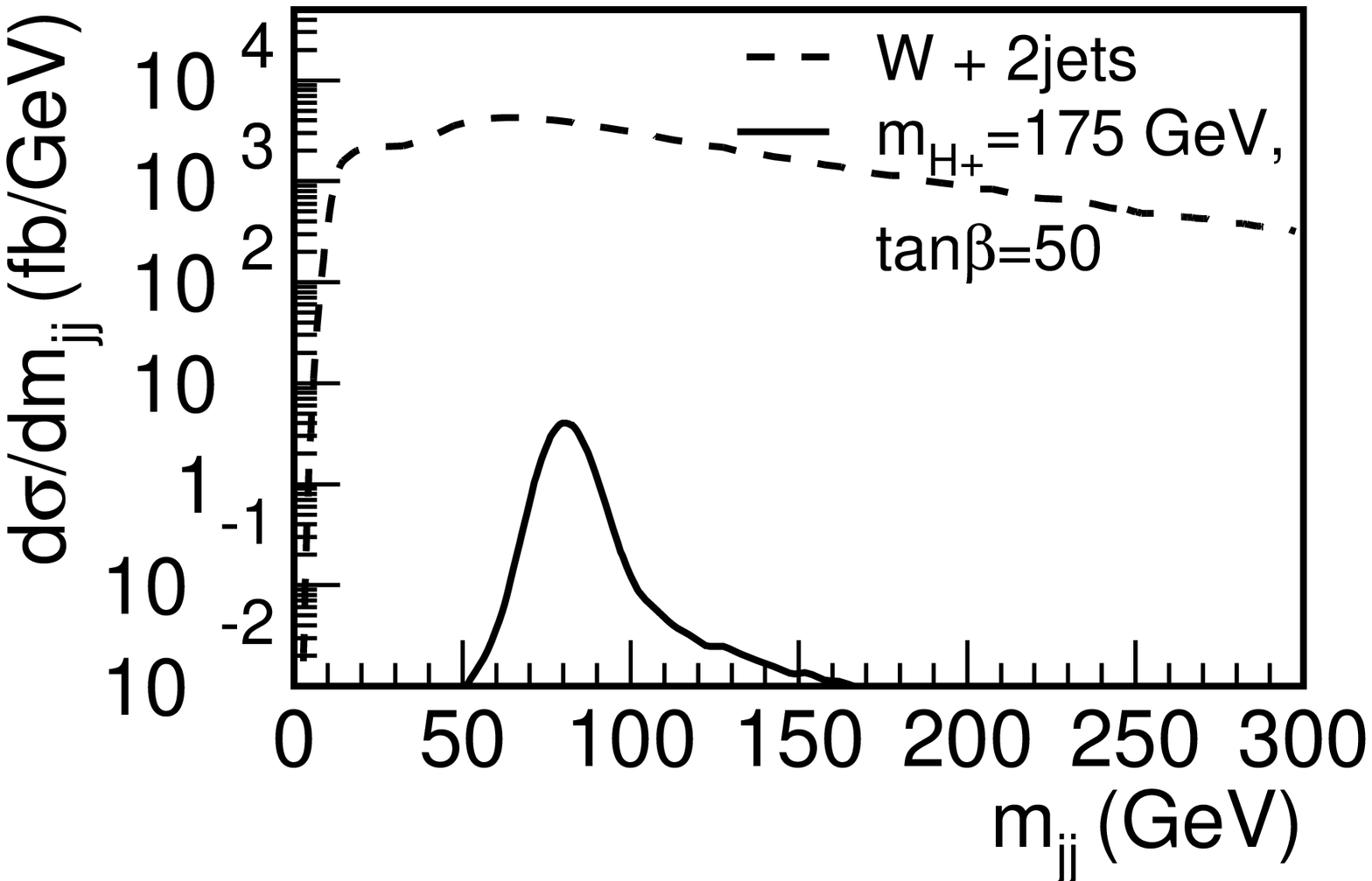} &
\includegraphics[width=7.5cm]{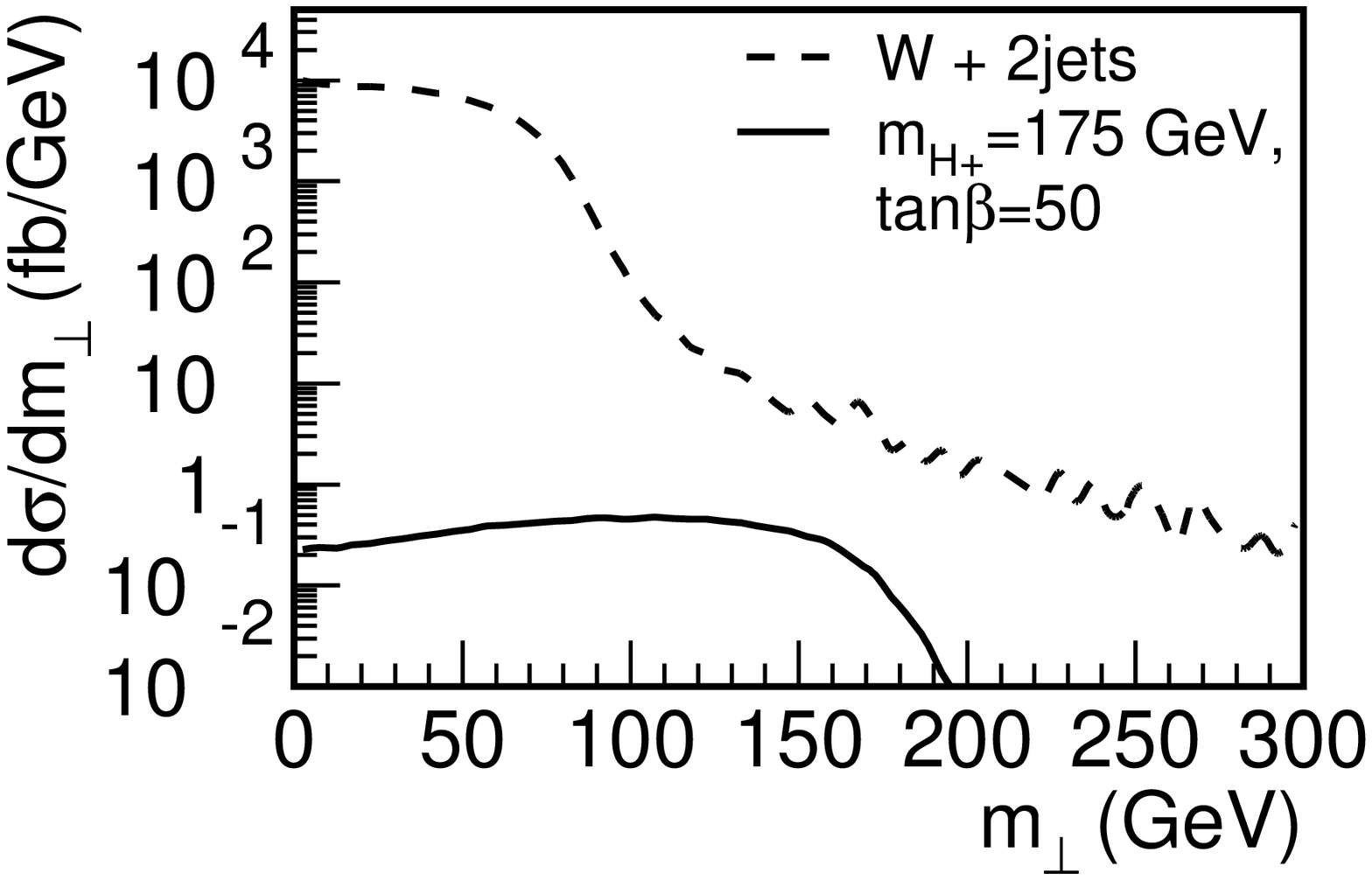} \\
\includegraphics[width=7.5cm]{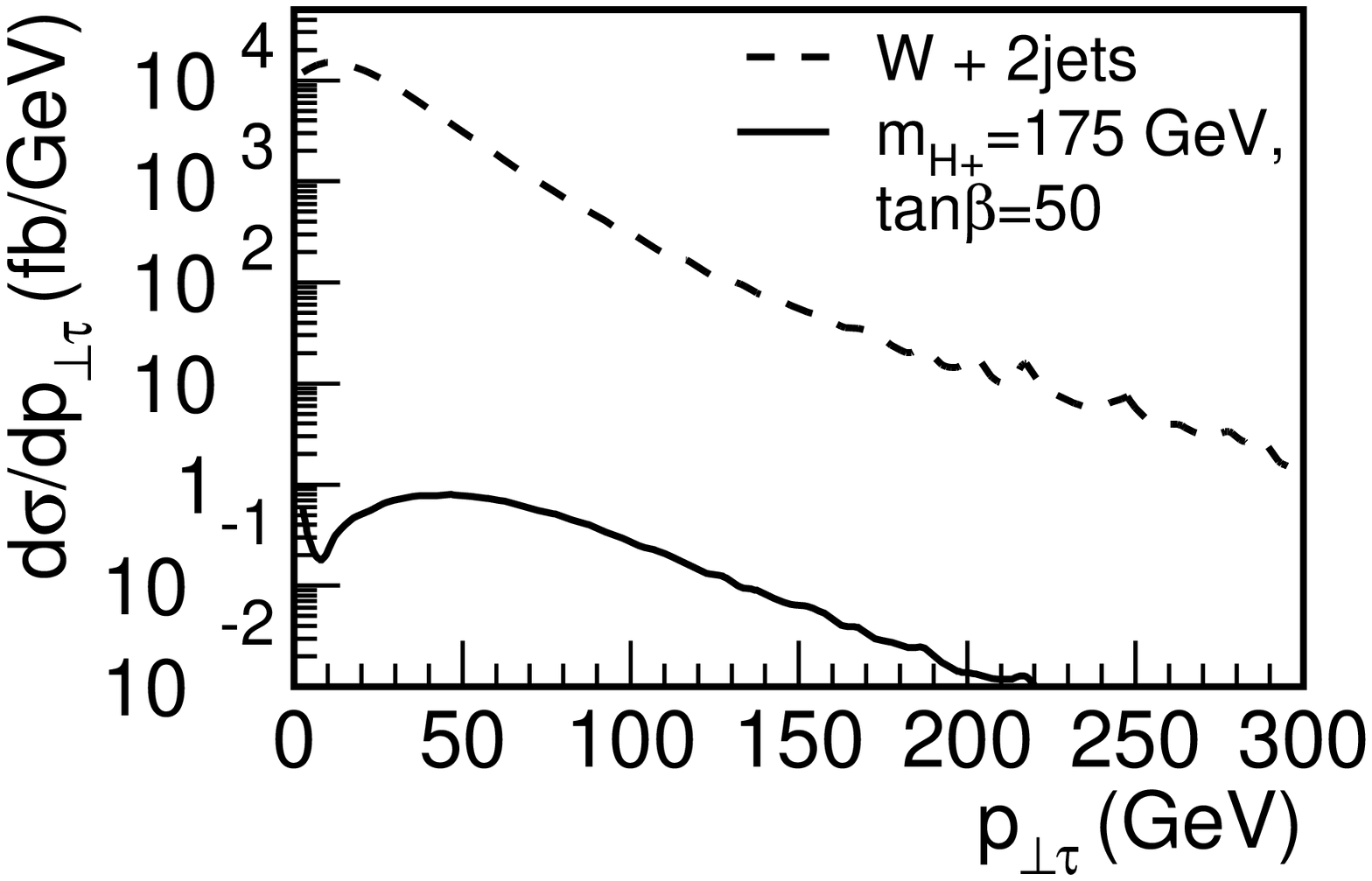} &
\includegraphics[width=7.5cm]{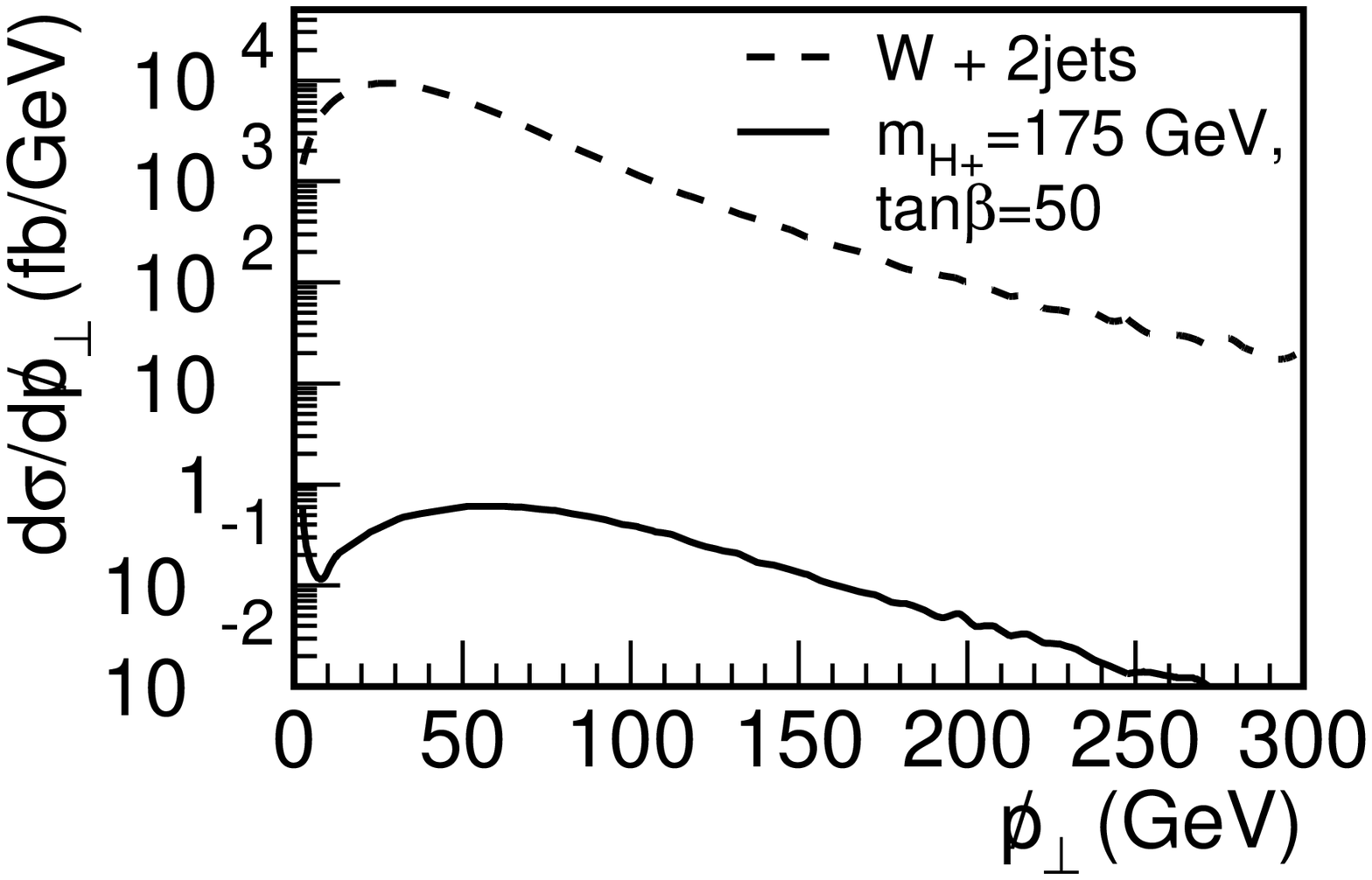} \\
\includegraphics[width=7.5cm]{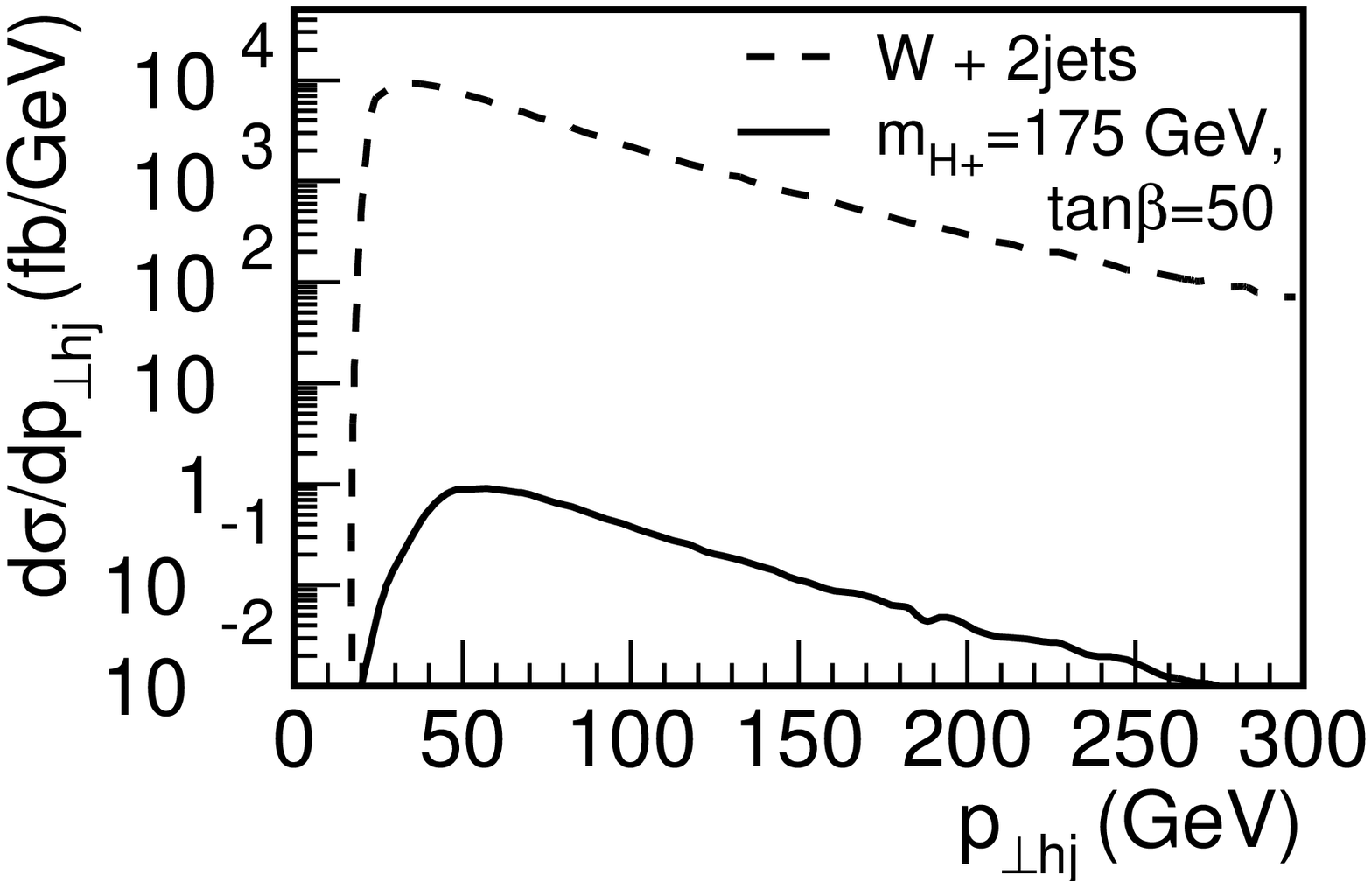} &
\includegraphics[width=7.5cm]{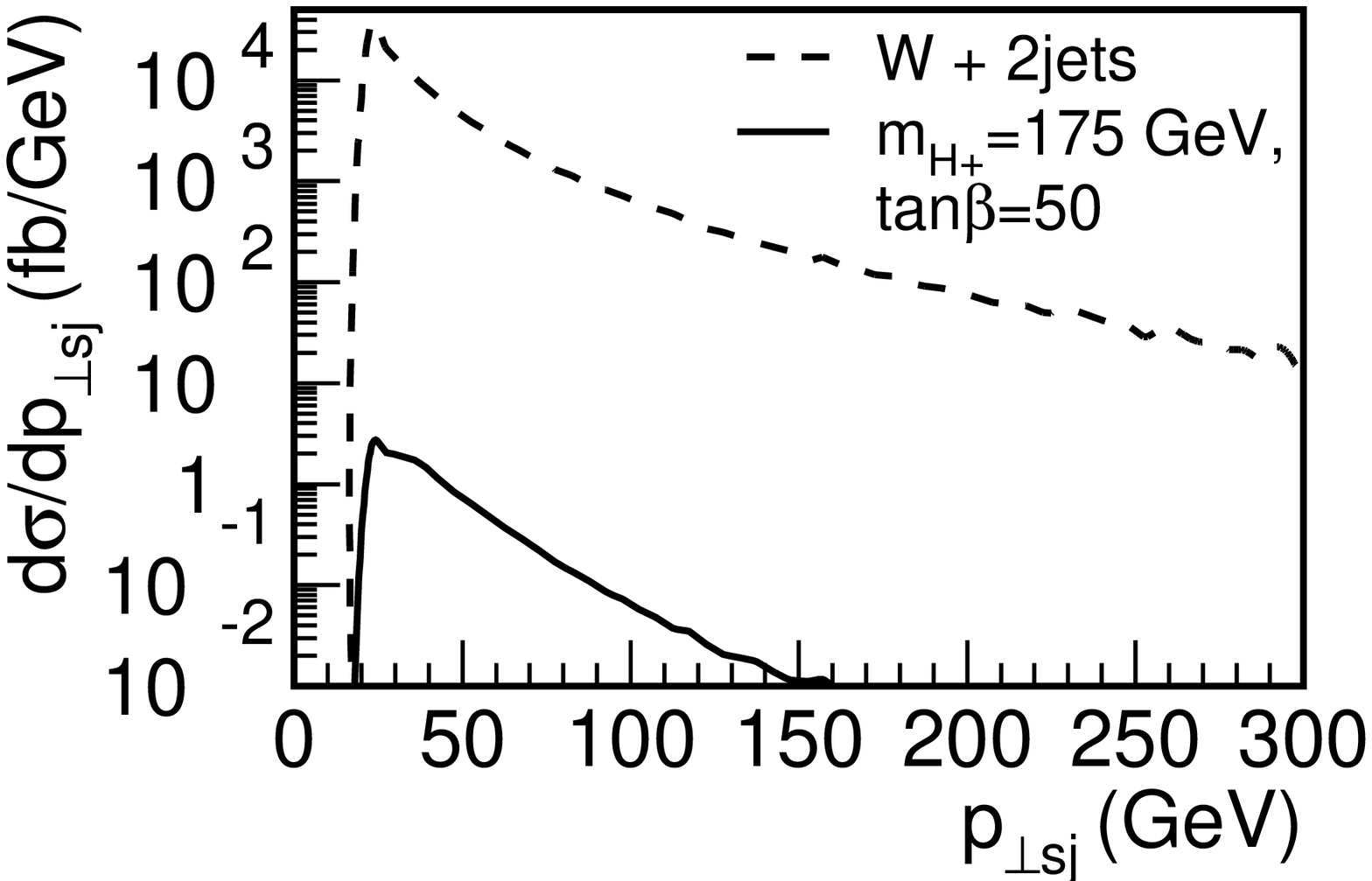} \\
\end{tabular}
\caption{Top row: $m_{jj}$ and $m_\perp$ distributions for signal (solid),
  and background (dashed). Middle row: $p_\perp$ distributions for
  $\tau_\mathrm{jet}$ and \mbox{$\!\not \!p$}. Bottom row: $p_\perp$ 
  distributions for the hard and soft light jet. 
  In all plots the basic cuts of table~\ref{tbcuts} have been
  applied.}\label{basiccuts}
\end{figure}

Our set of basic cuts, see table~\ref{tbcuts}, 
is defined according to the coverage and resolution
of a realistic detector. 
First we require that the pseudorapidity of the $\tau_\mathrm{jet}$ and the
light jets are in the range $-2.5<\eta<2.5$.
A further cut requires a minimal distance
$\Delta R_{jj}>0.4$ and $\Delta R_{\tau_\mathrm{jet} j}>0.5$
between two respective jets to allow their separation. 
Finally we require a minimum $p_{\perp}>20$ GeV  for all jets.
Figure~\ref{basiccuts} shows the resulting $m_{jj}$, $m_\perp$ distributions
and the $p_{\perp}$ distributions for all jets and the missing transverse 
momentum for the signal and background after these basic cuts.

Based on the shape of the signal and background in figure~\ref{basiccuts} we
define further cuts.
These cuts are used to reduce the background while keeping as much signal as
possible. 
It can be seen that a cut $70~\mathrm{GeV}<m_{jj}<90~\mathrm{GeV}$ will be very
efficient since the signal is peaked around the $W$ mass whereas the
background is almost flat.
A cut $m_\perp>100~\mathrm{GeV}$ will also remove a large part of the
background.
Increasing this cut leads to smaller $S/\sqrt{B}$ for $m_{H^\pm} = 175$~GeV,
yet might be useful for higher  $m_{H^\pm}$.
The signal peak has a soft upper edge, so for Higgs masses below 125 GeV it 
would be very
hard to see the signal since the background increases dramatically
in this region. 
To further reduce the background the cuts
$p_{\perp hj}>50$~GeV, $p_{\perp sj}>25$~GeV are defined.
In addition, we apply the cuts
$p_{\perp\tau_\mathrm{jet}}>50$~GeV and $\mbox{$\!\not \!p_{\perp}$}>50$~GeV,
similarly to \cite{Cao:2003tr}, 
in order to reduce the QCD background and the effects of detector 
misidentifications although we have not simulated these effects explicitly.
Since these cuts may be too soft
we will also show results for the harder cuts 
$p_{\perp\tau_\mathrm{jet}}>100$~GeV and $\mbox{$\!\not \!p_{\perp}$}>100$~GeV, which
have been used for example in~\cite{Assamagan:2002in}.

As mentioned above the $\tau$-jet properties of the signal and background 
are also different due to the difference in spin of the $H^\pm$ and $W$ bosons.
The relevant measure is the ratio, $R$, between the transverse momentum of
the leading charged $\pi$, $p_{\perp \pi}$, and the total transverse
momentum of the $\tau$ jet,
\begin{equation}
R =
{p_{\perp \pi} \over p_{\perp \tau_\mathrm{jet}}} .
\end{equation}
We assume that $p_{\perp \pi}$ is
measured in the tracker independently of the transverse momentum of the 
$\tau$-jet and in order to take into account the tracker performance
we apply  Gaussian smearing on $1/p_{\perp \pi}$ with
\begin{equation}
  \sigma(\frac{1}{p_{\perp \pi}})[\mathrm{TeV}^{-1}] = 
 \sqrt{0.52^2 + \frac{22^2}{(p_{\perp\pi}[\mathrm{GeV}])^2 \sin\theta_\pi}},
\end{equation}
where $\theta_\pi$ is the polar angle of the $\pi$ \cite{Richard}.

\begin{figure}
\centering
\begin{tabular}{cc}
\includegraphics[width=7.5cm]{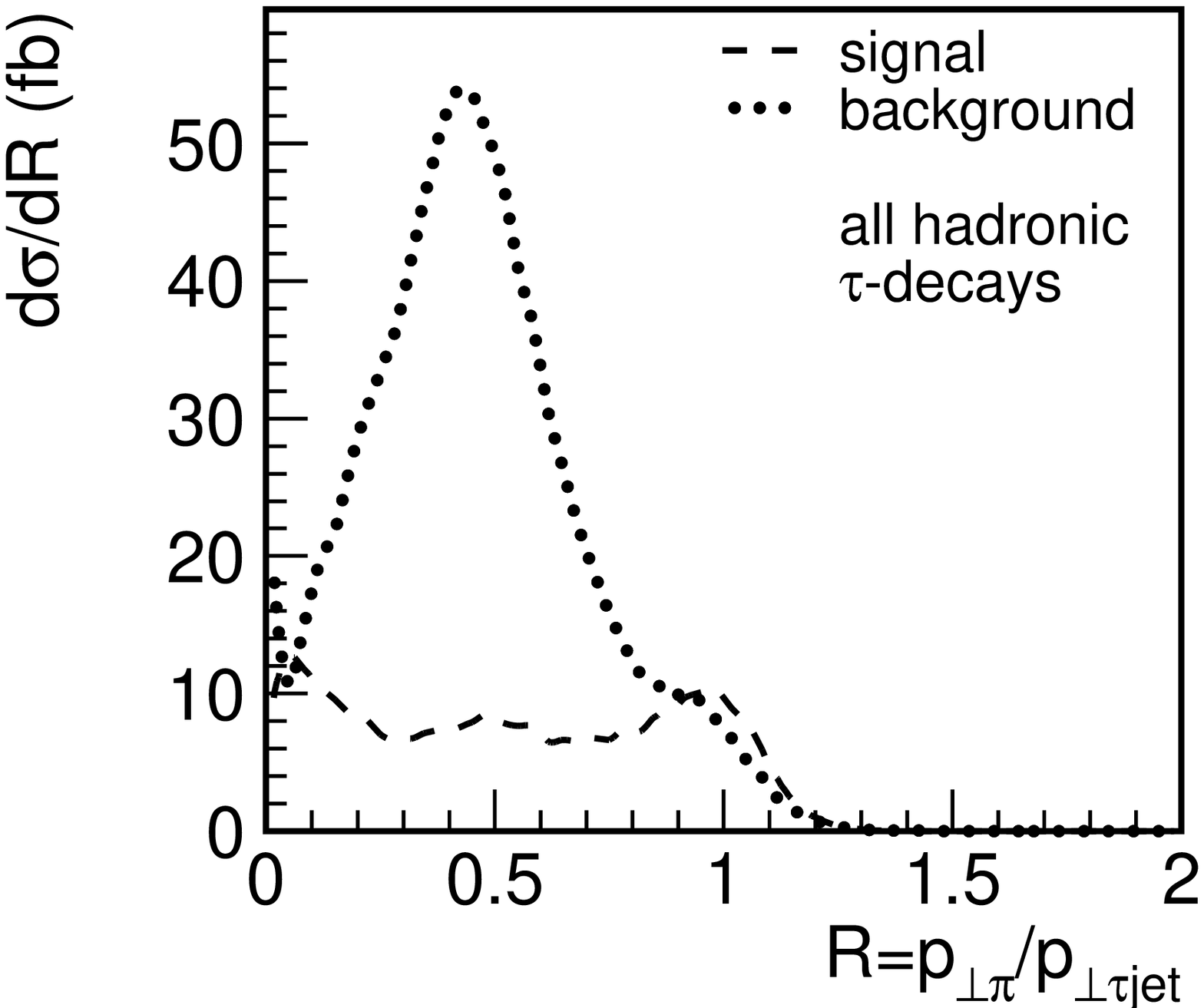} &
\includegraphics[width=7.5cm]{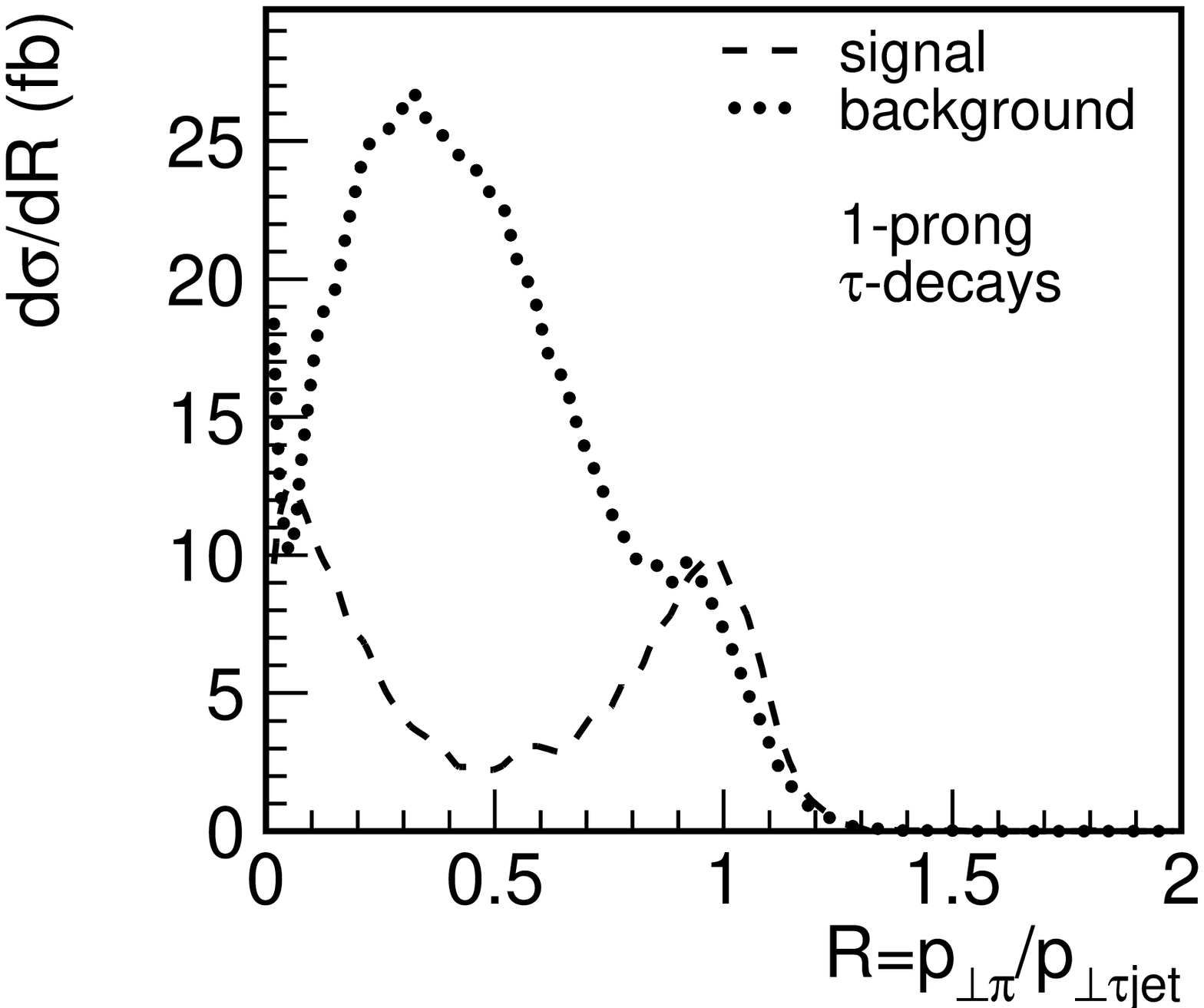} 
\end{tabular}
\caption{Comparison of $R={p_{\perp \pi} / p_{\perp \tau_\mathrm{jet}}}$ between signal (for
  $m_h^\mathrm{max}$ with $m_{H^\pm} = 175$ GeV  and $\tan\beta=50$)
  and background. The left plot is for all events and the right for
  1-prong events.
  The dashed (dotted) curve denotes signal (background).}\label{ptlpi_pttau}
\end{figure}

Having applied all the cuts outlined above, the resulting distribution of 
$R$ is shown in figure~\ref{ptlpi_pttau} both when keeping all
hadronic $\tau$-decays as well as only selecting 1-prong decays. 
As can be seen from the figure, the differences between the signal and
background is  clear in both cases.
Even so, we have found that applying cuts on $R$ in either of the two cases
does not lead to an improved overall significance $S/\sqrt{B}$ compared to 
the case when all hadronic $\tau$-decays are kept and no cut on $R$ is applied.
(When selecting only 1-prong events a cut $R>0.8$ turns out to be
advantageous, but the overall significance still drops compared to keeping all
hadronic $\tau$-decays.) 
At the same time it should be kept in mind that these
conclusions may change when making a complete simulation of the final state
including parton showers and hadronisation as well as a full detector simulation.
It may also be possible to enhance the 
use of $\tau$-polarization by considering the 3-prong decays separately
and looking at the sum of the transverse momenta of the like-sign pair instead
of that of the leading charged pion~\cite{Guchait:2006jp}.

\begin{table}
\centering
\begin{tabular}{cc}\hline\hline
Basic cuts & Additional cuts [all in GeV]
 \\\hline
$|\eta_{\tau_\mathrm{jet}}|<2.5$
& $p_{\perp\tau_\mathrm{jet}}>50$, $\mbox{$\!\not \!p_{\perp}$}>50$
 \\
$|\eta_j|<2.5$
& $70$ $<m_{jj}<90$ 
 \\
$\Delta R_{jj}>0.4$
& $m_\perp>100$  
 \\
$\Delta R_{\tau_\mathrm{jet} j}>0.5$
& $p_{\perp hj}>50$, $p_{\perp sj}>25$  
 \\
$p_{\perp jet}>20$ GeV & 
 \\\hline \hline
\end{tabular}
\caption{Set of basic cuts which define a
  signal region that corresponds to the sensitive region of a real detector
  and additional cuts which reduce the background according to the 
  distributions in figure
  \ref{basiccuts} as well as
  suppress QCD background and detector misidentifications.}\label{tbcuts} 
\end{table}

In table~\ref{tbcuts} all basic and additional cuts are summarized and
in table~\ref{cuteffect} it is shown how applying the different additional
cuts one after the other affects the integrated cross-section.
The signal is given for the $m_h^\mathrm{max}$ scenario in table
\ref{maxmixpar} with a charged Higgs mass of 175 as well as 400 GeV 
and $\tan\beta=50$. 
We have used an integrated luminosity of 300 fb$^{-1}$ and a $\tau$
detection efficiency of 30\% to calculate $S/\sqrt{B}$. 
Finally we note that, 
when the charged Higgs mass is large it can be advantageous to use 
the harder set of cuts $p_{\perp\tau_\mathrm{jet}}>100$~GeV and
 $\mbox{$\!\not \!p_{\perp}$}>100$~GeV as illustrated below.
In order to get a rough estimate of how important higher order corrections 
can be on the resulting significances we consider the following worst case 
scenario. Assuming the total uncertainty in the signal as well as background 
to be a factor $1.5$, then the significances in table~\ref{cuteffect} would in 
the worst case be reduced with a factor $1.8$. 

\begin{table}
\centering
\begin{tabular}{c|c|ccc|ccc} \hline \hline
 & & \multicolumn{3}{c}{ $m_{H^\pm}=175$ GeV} &  
 \multicolumn{3}{c}{$m_{H^\pm}=400$ GeV}  \\
Cut [all in GeV] & $\sigma_{\rm b}$ (fb) & $\sigma_{\rm s}$ (fb) & 
$S$ & $S/\sqrt{B}$& $\sigma_{\rm s}$ (fb) & $S$ & $S/\sqrt{B}$ \\ \hline
Basic cuts &
 560000 & 55 & 4900  & 0.7 & 3.3 & 300 & 0.04 \\
$p_{\perp\tau_\mathrm{jet}}>50$ , $\mbox{$\!\not \!p_{\perp}$}>50$ &
 22000 & 25 & 2200 &  1.6 & 2.7 & 240 & 0.2 \\
$70 <m_{jj}<90$  &
 1700 & 21 & 1900 &  5 & 2.2 & 200 & 0.5 \\
$m_\perp>100$  &
 77 & 15 & 1400 &  16 & 2.1 & 190 & 2.3 \\
$p_{\perp hj}>50$  , $p_{\perp sj}>25$  &
 28 & 9.3 & 840 &  17 & 1.5 & 135 & 2.6 \\ 
 \hline \hline
\end{tabular}
\caption{The effect of the different cuts on the integrated cross-section for
  background  ($\sigma_{\rm b}$) and signal  ($\sigma_{\rm s}$) in the
  $m_h^\mathrm{max}$ scenario
  with $m_{H^\pm}=175$ and 400 GeV for $\tan\beta=50$ (see
  table~\ref{maxmixpar})
  as well as the number of signal events $S$ and the significance $S/\sqrt{B}$
  assuming an integrated luminosity of 
  300 fb$^{-1}$ and a $\tau$ detection efficiency of 30\% .
}\label{cuteffect} 
\end{table}

\section{Results}
In this section we present the results of our analysis in the MSSM with real 
and complex parameters as well as in special scenarios with large mass 
splittings giving resonant enhancement of the signal.

\subsection{MSSM with real parameters}

Unless otherwise noted we use in the following
a standard maximal mixing scenario, $m_h^\mathrm{max}$, with
$X_t=2M_{SUSY}$  as defined in \cite{Carena:2002qg}.
All SUSY parameters needed as input to calculate the Higgs masses and
mixing matrix with {\sc FeynHiggs} are given in table~\ref{maxmixpar}.
For the electroweak parameters, $m_t=178$ GeV, $m_b=4.7$ GeV,
$m_W=80.426$ GeV, and $m_Z=91.187$ GeV have been used.
$G_F$ has been calculated from the running $\alpha_{EW}$ as
$G_F={\pi\alpha_{EW} / (\sqrt{2}\sin^2\theta_W m_W^2)}$. 
We have checked that the impact of using a lower $m_t=172$ GeV on the
signal rates and distributions is small compared to the uncertainties
from higher order effects and can safely be neglected. This also applies 
to the branching ratio $Br (H^+ \to \tau^+ \nu_\tau)$, which in the case of
$m_{H^\pm}=400$ GeV is reduced from 22.7 to 22.0 \%.

\begin{figure}
\centering
\begin{tabular}{cc}
\includegraphics[width=7.5cm]{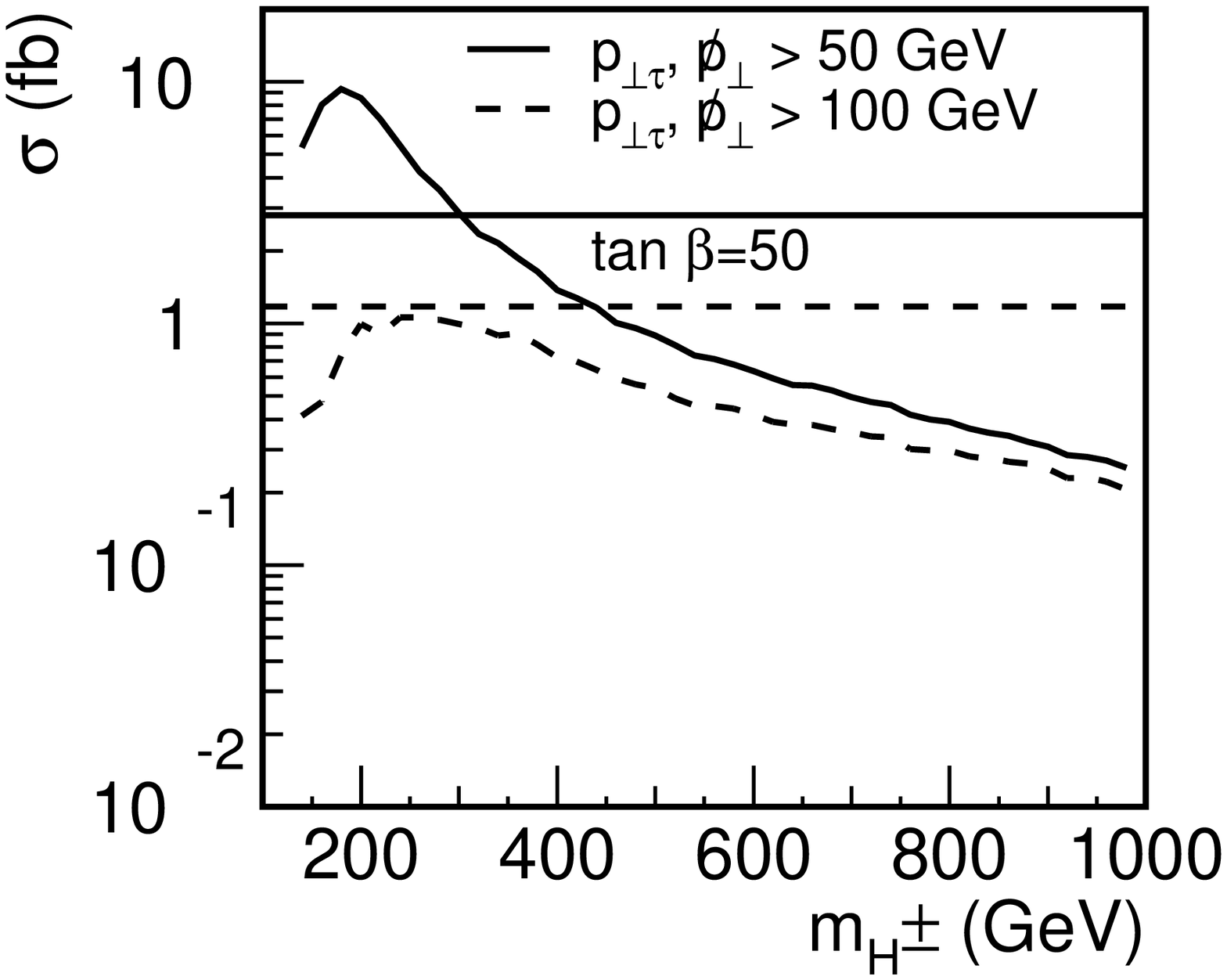} &
\includegraphics[width=7.5cm]{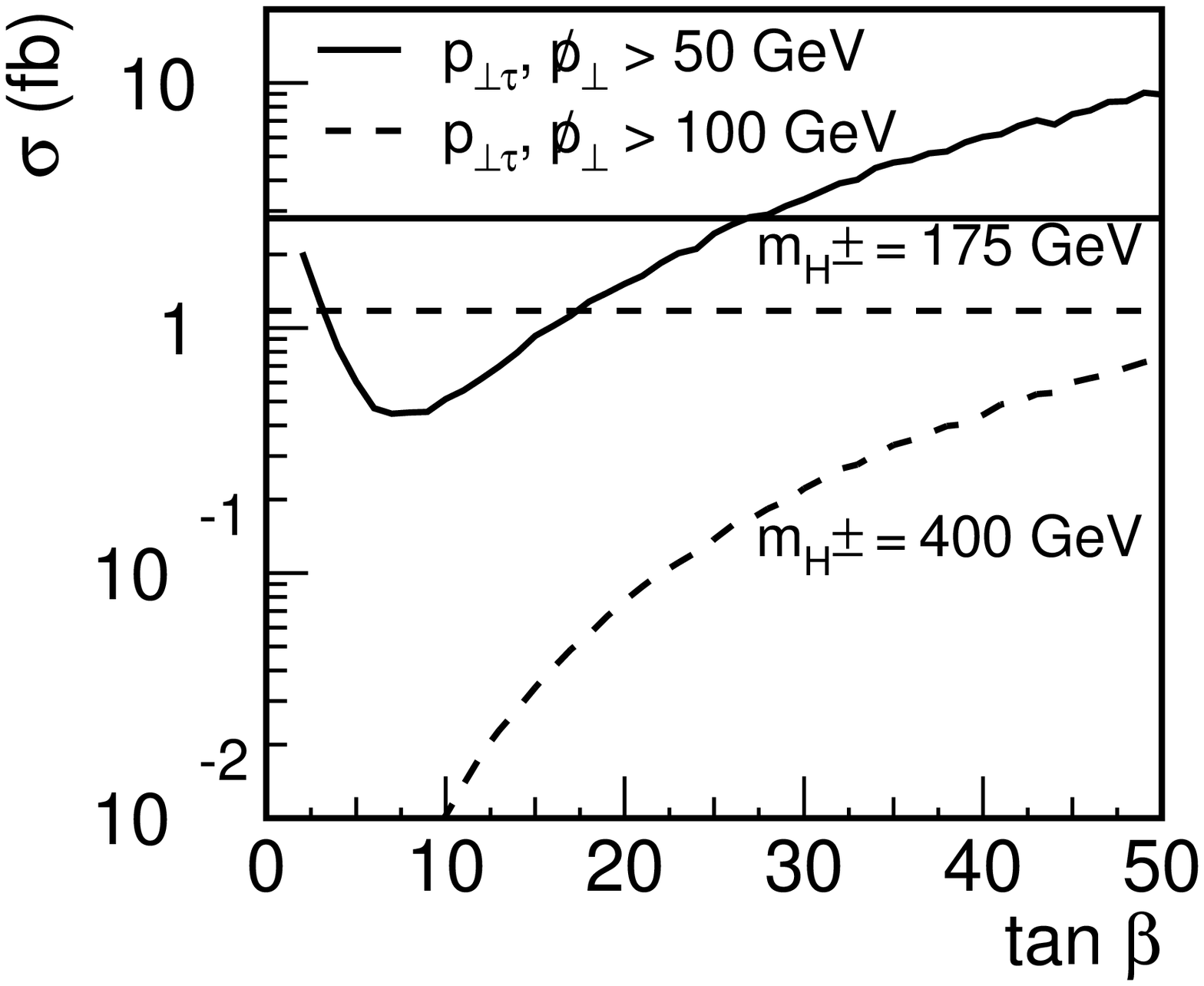}\\
\end{tabular}
\caption{$H^\pm$ mass and $\tan\beta$ dependence of the integrated
  cross-section in the $m_h^\mathrm{max}$ scenario. 
  Solid curves are with all cuts
  of table~\ref{tbcuts}, and dashed curves are with the harder  cuts 
  $p_{\perp\tau_\mathrm{jet}}>100$ GeV and $\mbox{$\!\not \!p_{\perp}$}>100$ GeV.
  The horizontal lines
  correspond to $\frac{S}{\sqrt{B}}=5$.
}\label{cutsigma_mhctanbeta} 
\end{figure}

The mass and $\tan\beta$ dependence of the cross-section after all cuts 
of table~\ref{tbcuts} are shown in figure~\ref{cutsigma_mhctanbeta}
as solid curves whereas dashed curves denote the cross-section
for the harder cuts 
$p_{\perp\tau_\mathrm{jet}}>100$ GeV and 
$\mbox{$\!\not \!p_{\perp}$}>100$ GeV.
The left plot is for $\tan \beta=50$ whereas
in the right plot we have
used $m_{H^\pm}=175$~GeV
(400~GeV) for the solid (dashed) line.
The horizontal lines indicate the cross-section needed for
$\frac{S}{\sqrt{B}}=5$.
Using this criterion to define a detectable signal 
we see that this would indeed be the
case for $\tan\beta \gtrsim 30$ if $m_{H^\pm}=175$~GeV and for 
$150~\textrm{GeV} \lesssim m_{H^\pm} \lesssim 300 $~GeV  if $\tan\beta =50$
with the softer cuts $p_{\perp\tau_\mathrm{jet}}>50$ GeV and 
$\mbox{$\!\not \!p_{\perp}$}>50$ GeV, whereas with the harder cuts
$\tan\beta$ has to be larger than at least 50 in order to have a detectable
signal.

\begin{figure}
\centering
\includegraphics[width=7.5cm]{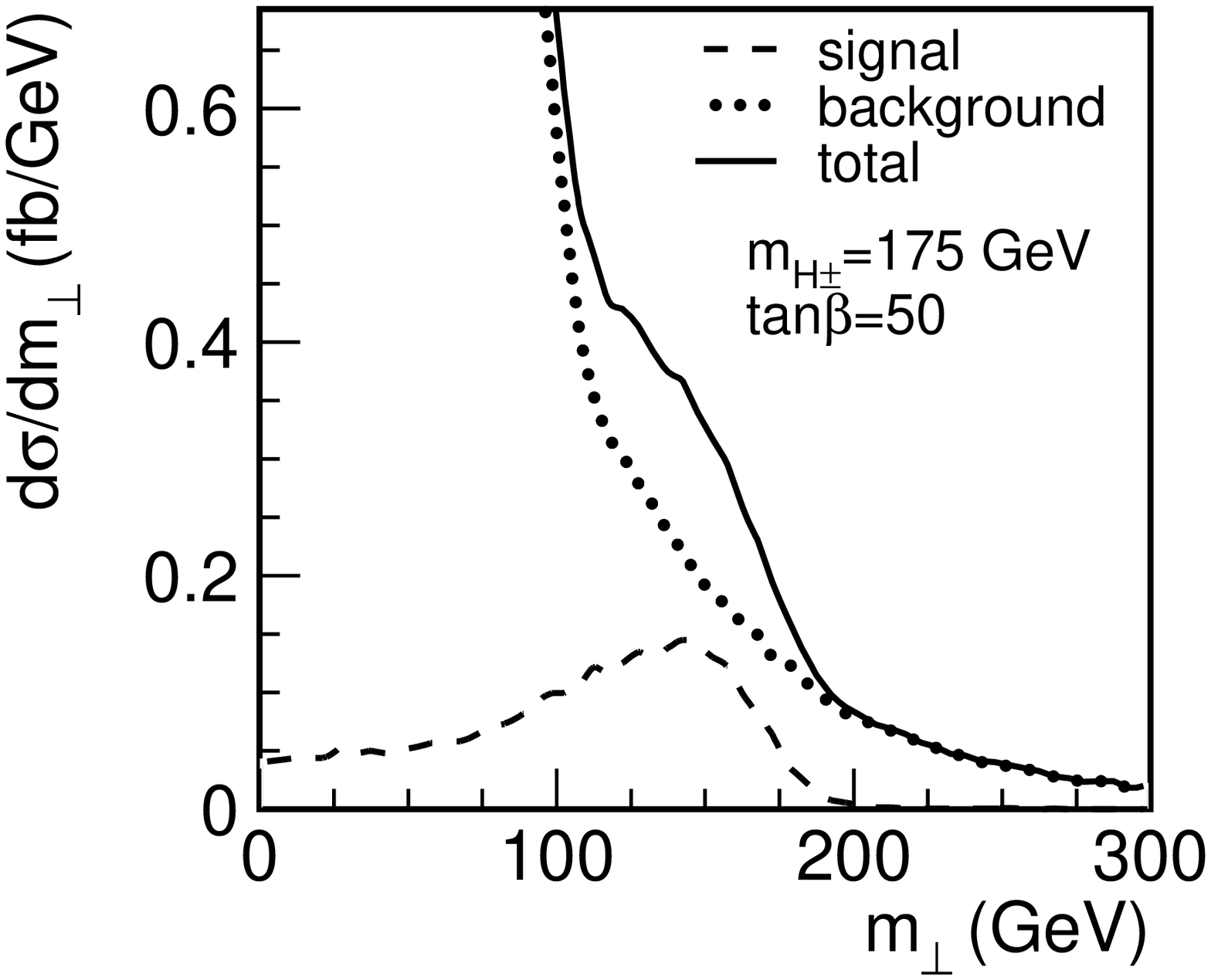}
\includegraphics[width=7.5cm]{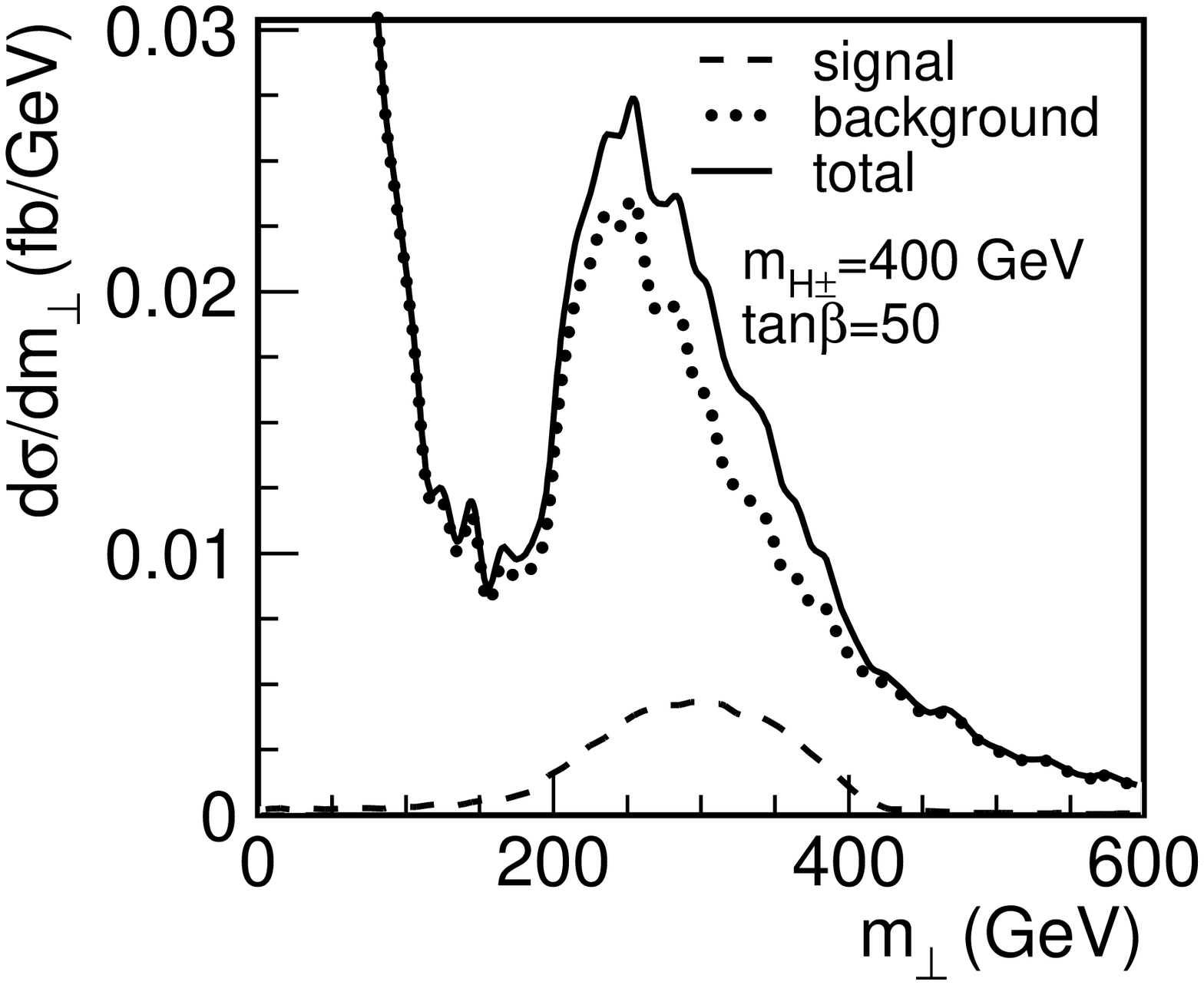}
\caption{Comparison of the $m_\perp$ distribution between the signal in the
  $m_h^\mathrm{max}$ scenario with $\tan\beta=50$ and $m_{H^\pm} = 175$ GeV
  (left) as well as $m_{H^\pm} = 400$ GeV (right) together with the respective 
  backgrounds with all cuts of table~\ref{tbcuts}
  (for the high mass the harder cuts 
  $p_{\perp\tau_\mathrm{jet}} > 100$~GeV and 
  $\mbox{$\!\not \!p_{\perp}$} > 100$~GeV are
  used). The dashed curve denotes the signal and the dotted one 
  the background whereas the solid curve is the sum of the two. }\label{m175} 
\end{figure}

Figure~\ref{m175} shows the resulting 
$m_\perp$ distribution for a charged Higgs mass of 175 GeV as well as
400 GeV in the case $\tan\beta =50$ compared to the background after 
all cuts in table~\ref{cuteffect} have been applied.
In the high mass case the harder cuts
$p_{\perp\tau_\mathrm{jet}} > 100$~GeV and $\mbox{$\!\not \!p_{\perp}$} > 100$~GeV
are used but all the other cuts are the same.
For an integrated luminosity of 300 fb$^{-1}$ and a $\tau$ detection efficiency
of 30\% we get $S/\sqrt{B}=17$ for $m_{H^\pm} = 175$ GeV and 
  $S/\sqrt{B}=3.2$ for $m_{H^\pm} = 400$ GeV. (The latter result is slightly
  better than what would have been obtained with the softer cuts
$p_{\perp\tau_\mathrm{jet}} > 50$~GeV and 
$\mbox{$\!\not \!p_{\perp}$} > 50$~GeV.)
Here, and in all following
calculations of $S/\sqrt{B}$, we have used the cut $m_\perp>100$ GeV 
to get $S$ and $B$. In the high mass case $S/\sqrt{B}$
could in principle be improved by imposing a harder cut on $m_\perp$, 
as can be seen from the figure, but not much.
Also the possible use of upper cuts $m_\perp < 200$~GeV ($m_\perp < 500$~GeV)
for $m_{H^\pm} = 175$~GeV ($m_{H^\pm} = 400$~GeV)
only marginally improves $S/\sqrt{B}$ from 17 (3.2) to 19 (3.3).
In the same figure we also see that the harder cuts create a fake peak in the
background which could make it more difficult to tell if there is a signal. 
This fake peak appears since the $\tau$ and $\nu_\tau$ are mostly produced 
back to back in the $W+2$ jets process. Using the harder cuts 
$p_{\perp\tau_\mathrm{jet}} > 100$~GeV and 
$\mbox{$\!\not \!p_{\perp}$} > 100$~GeV,
to reduce  QCD background and detector 
misidentifications, the significance 
for $m_{H^\pm} = 175$ GeV and $\tan\beta =50$ is reduced to
 $S/\sqrt{B}=3.1$. However, in this case using a upper cut 
 $m_\perp < 200$~GeV is beneficial leading to a significance of 
 $S/\sqrt{B}=6.4$.
This means that if the softer cuts
$p_{\perp\tau_\mathrm{jet}} > 50$~GeV and $\mbox{$\!\not \!p_{\perp}$} > 50$~GeV
are sufficient to suppress the QCD background these are
clearly to be preferred.

\subsection{MSSM with complex parameters}

\begin{figure}
\centering
\begin{tabular}{ccc}
\includegraphics[width=5cm]{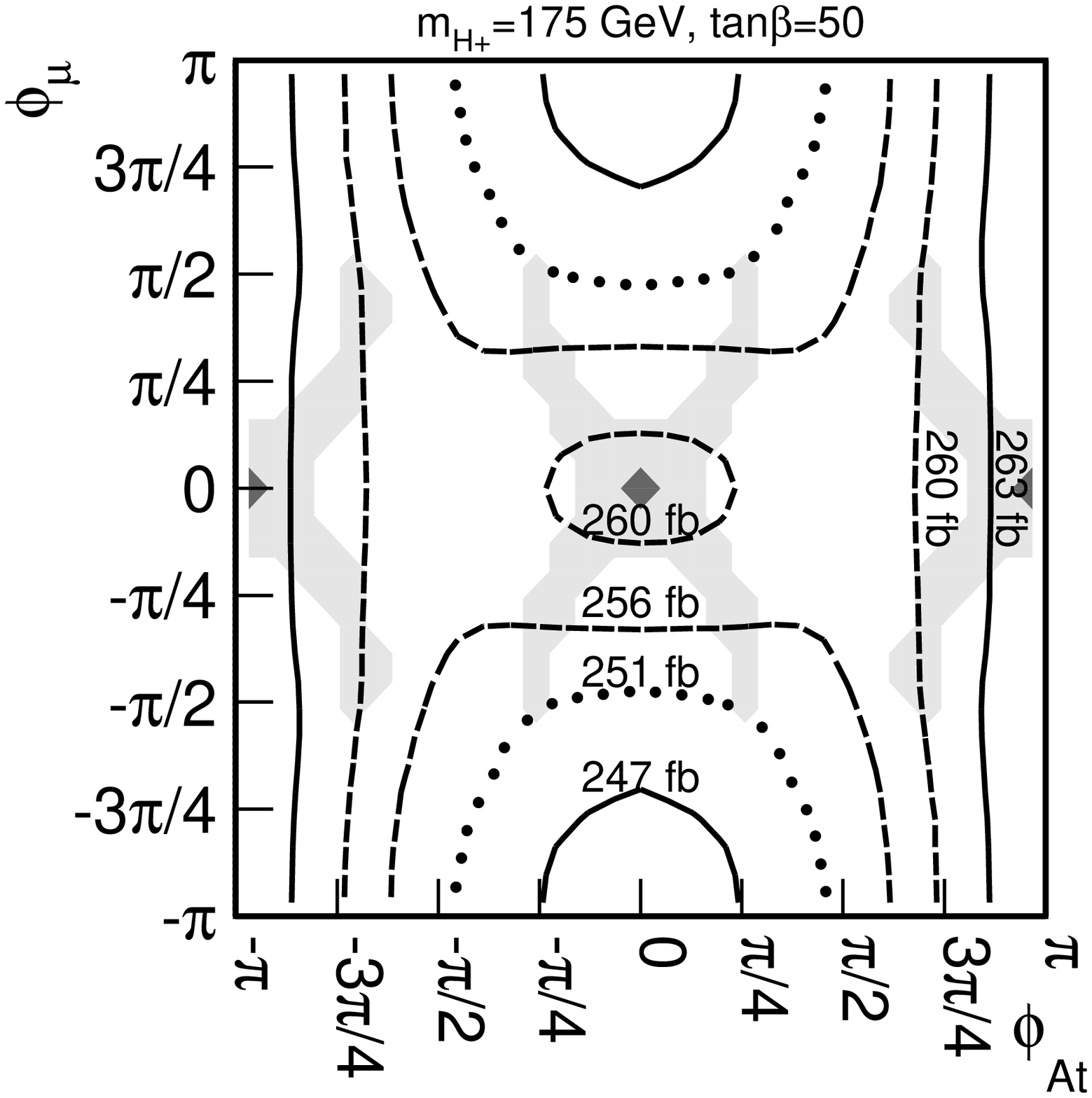} &
\includegraphics[width=5cm]{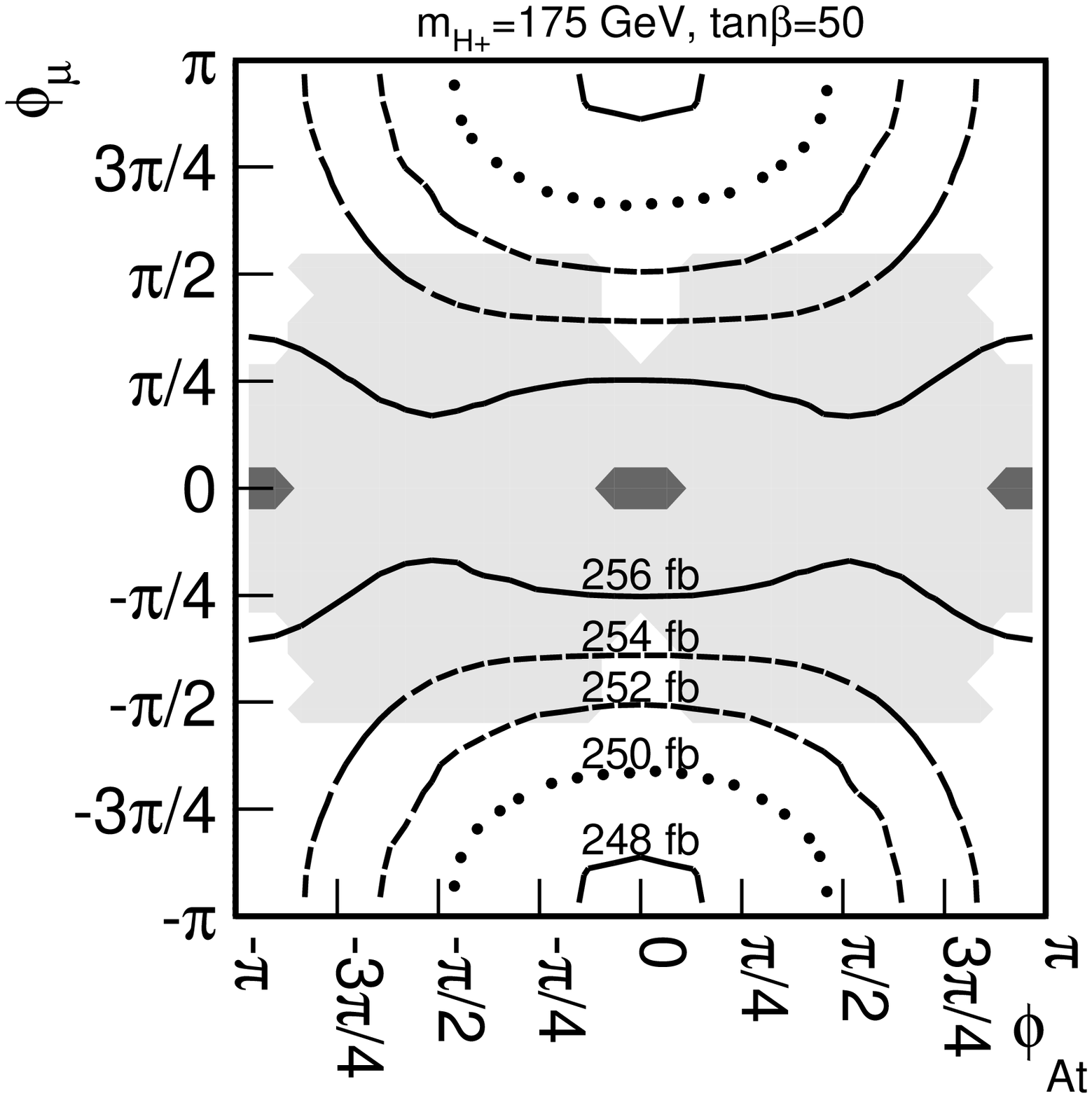} &
\includegraphics[width=5cm]{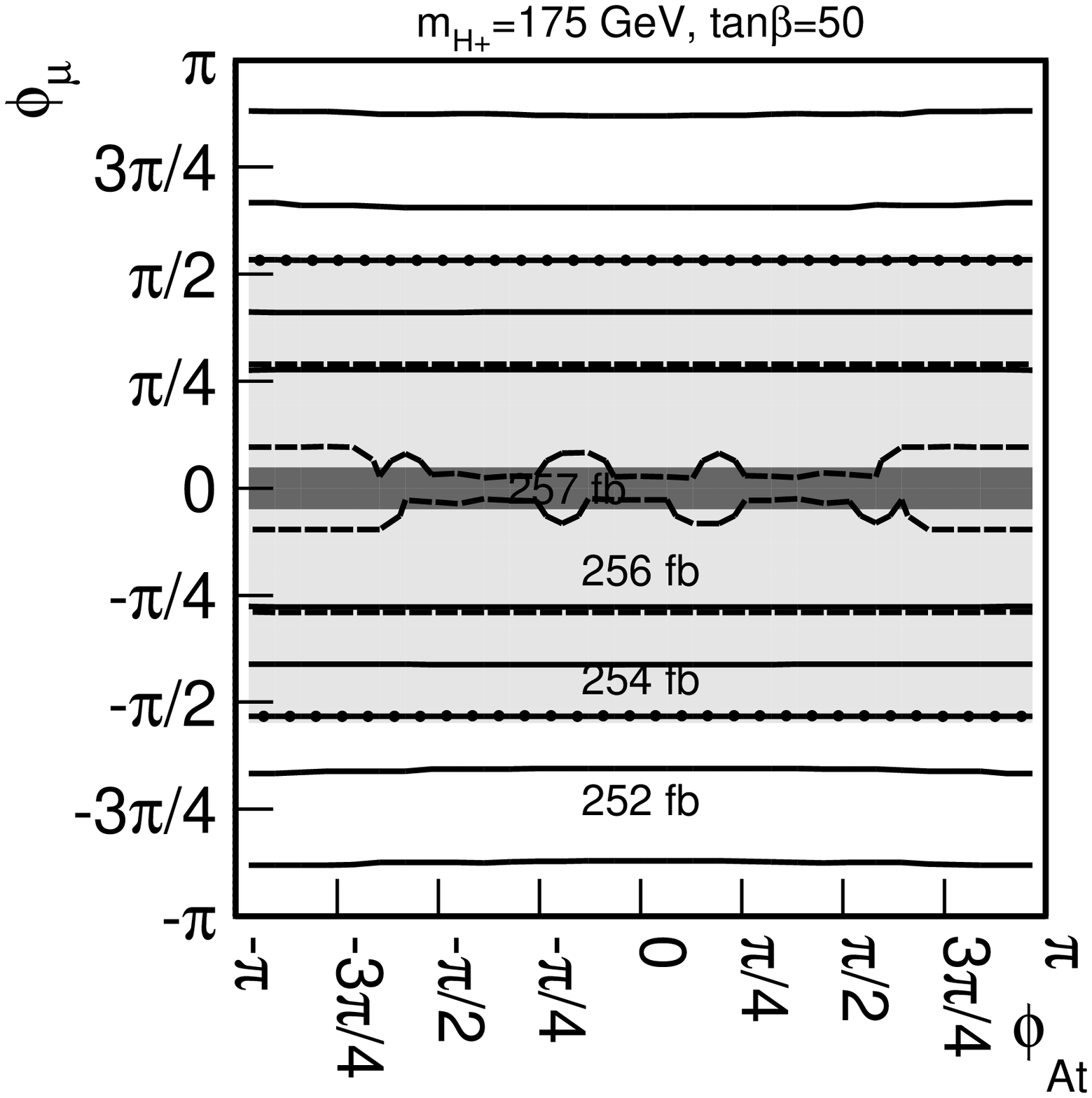}
\end{tabular}
\caption{Total cross-section as a function of $\phi_\mu$ and $\phi_{A_t}$ in
  the maximal mixing scenario (left), the less mixing scenario (middle) 
  and the no mixing scenario (right). 
  The light shaded areas are in
  agreement with the constraints from $a_\mu$ and $\delta\rho_0$ and 
  the dark shaded ones 
  are also in agreement with the constraints from EDMs.}\label{fig:phases}
\end{figure}

In the general MSSM many parameters can be complex. 
However, our signal process, analyzed as described in section 2,
is only affected by CP violation in the neutral Higgs sector because
of the neutral Higgs bosons exchanged in the $s$-channel.
The leading contributions to the CP violation in the neutral Higgs
sector arise from loops of the scalar top and (to a lesser extend)
of the scalar bottom sector 
where the possibly complex Higgs/higgsino mass parameter $\mu$
and the trilinear scalar couplings  $A_t$ and $A_b$ are dominant.
Furthermore, in constrained MSSM scenarios implying universality
conditions at the GUT scale and rotating away all unphysical phases
with help of U(1) symmetries of the theory, only two phases remain,
the phase of $\mu$ and a common phase for the trilinear couplings
\cite{Pilaftsis:1998dd,Pilaftsis:1999qt}.
Hence, we concentrate in the following on 
the phases $\phi_\mu$ and $\phi_{A_t}$ of $\mu$ and $A_t$, respectively, which
have the largest effect on the neutral Higgs
sector and thus possibly affect our signal, assuming
$\phi_{A_b}=\phi_{A_\tau}=\phi_{A_t}$.
We have varied $\phi_\mu$ and
$\phi_{A_t}$ independently in the range $-\pi < \phi < \pi$ in order
to investigate the phase dependence of our signal.
However, in the maximal mixing scenario as well as in scenarios with
less and no mixing in the third generation squark sector,
see table~\ref{maxmixpar}, 
we find only small ($\sim 5 \%$)
$\phi_\mu$ and $\phi_{A_t}$ dependencies of the total cross-section as can be
seen in figure~\ref{fig:phases}.

The small effects of the phases in these MSSM scenarios are basically 
due to the neutral Higgs bosons having quite similar masses 
 ($m_{H_i}^2 - m_{H_j}^2 \ll (m_{H^\pm}+m_W)^2 - m_{H_i}^2$)
and small widths with
the latter also being true for the charged Higgs boson.
Bearing in mind that $s\gtrsim (m_{H^\pm}+m_W)^2$, the propagators, 
$S_{H_i}=1 / (s - m_{H_i}^2+im_{H_i}\Gamma_{H_i})$, in
Eq.~(\ref{eq:bbhpwm}) and (\ref{eq:bbhmwp})
are all of similar size and approximately real. 
Therefore, as a first approximation, they can be put in
front of the sum over the different Higgs bosons.
The remaining sums of coupling factors then reduce to simple numbers, 
the first to $2\sin^2\beta$ and the second to $-2\sin\beta$, 
irrespective of the entries in the Higgs mixing matrix, $O_{ji}$.
Thus, unless we have 
large differences in the masses of the neutral Higgs bosons,
the effect from the phases via the couplings is small.
The  phase dependence also 
enters indirectly via the Higgs masses, which typically
have a large dependence on the phases. 
It turns out that in the scenarios we are considering 
the total cross-section has 
an almost linear dependence on the mass of the heaviest neutral Higgs boson and
the phase dependence from the masses is much larger than that 
from the couplings.

Figure~\ref{fig:phases} also shows the allowed regions from the following 
precision measurements, the 
anomalous magnetic moment of the muon, $a_\mu$, the $\rho$-parameter, 
$\delta\rho_0$ and the electric dipole moments (EDMs) of $e$, $n$ and Hg. 
The experimental value of $a_\mu$ from \cite{Bennett:2004pv} is
\begin{equation}
a_\mu^\mathrm{exp.} = 11\, 659\, 208.0 \pm 5.8 \times 10^{-10}
\end{equation}
and the theoretical value determined using $e^+e^-$ data from
\cite{Davier:2004gb} is 
\begin{equation}
a_\mu^\mathrm{SM} = 11\, 659\, 184.1 \pm 8.0 \times 10^{-10}.
\end{equation}
Based on these values we use the following $2\sigma$-range for the difference
\begin{equation}
-3.7 \times 10^{-10} < \Delta a_\mu < 51.5 \times 10^{-10}.
\end{equation}
The experimentally allowed range of $\delta\rho_0$ is \cite{Eidelman:2004wy}
\begin{equation}
-0.0010 < \delta\rho_0 < 0.0025,
\end{equation}
whereas we use the following upper limits 
on the EDMs 
\cite{Regan:2002ta,Harris:1999jx,Romalis:2000mg}
\begin{equation}
|d_e|<1.6\cdot 10^{-27},
\end{equation}
\begin{equation}
|d_n|<6.3\cdot 10^{-26},
\end{equation}
\begin{equation}
|d_\mathrm{Hg}|<2.1\cdot 10^{-28}.
\end{equation}
In order to calculate $a_\mu$, $\delta\rho_0$, $|d_e|$, $|d_n|$ and
$|d_\mathrm{Hg}|$ in our scenarios and to analyze their dependence on the 
phases of $\mu$ and $A_t$ we have used {\sc FeynHiggs} 2.2.10, where, however, 
it is important to keep in mind that the
EDM routines are not yet fully tested.
As can be seen from figure~\ref{fig:phases} the restrictions from the precision
measurements are very severe. 
The bound on $a_\mu$ effectively removes the region with 
$|\phi_\mu| \gtrsim \pi/2$
and in the remaining region only 
the light shaded areas are consistent with the $\delta\rho_0$ bound.
Finally the darker shaded areas are those consistent also with the EDMs.
Hence only small variations from phase zero or $\pi$ are allowed.

The constraints from the EDMs are very strong in constrained MSSM scenarios, 
allowing only small values of the phases, especially
of $\phi_\mu$. However, in unconstrained SUSY they are rather
model dependent, permitting in general larger phases.
For example, due to cancellations between different SUSY contributions to
the EDMs
or in SUSY models with heavy sfermions in the first two generations
larger phases may be allowed
\cite{Ibrahim:1998je,Ibrahim:1999af,Bartl:1999bc,Barger:2001nu,Abel:2001vy,Choi:2004rf,Pospelov:2005pr,Olive:2005ru,Abel:2005er}.
Recently it has been pointed out that for large trilinear scalar couplings $A$,
phases $\phi_\mu\sim O(1)$ can be compatible with
the bounds on $d_e$, $d_n$ and $d_\mathrm{Hg}$~\cite{YaserAyazi:2006zw}.
Furthermore the restrictions on the phases may also disappear if
lepton flavour violating terms in the MSSM Lagrangian are included
\cite{Bartl:2003ju}.
In conclusion this means that 
large phases cannot be ruled out and therefore we analyze the full range
of the phases to determine possible effects.

To see the dependence of the allowed regions on the chosen SUSY scenario
we show in
figure~\ref{fig:phases} also the results for the less mixing and no
mixing scenarios.
In the less mixing scenario the constraints from $\delta\rho_0$ are less
severe but the EDMs still give hard constraints whereas for the no mixing
scenario we get a complete band around $\phi_\mu=0$. Finally we have also
checked that similar results are obtained for $\tan\beta=10$.

In the MSSM scenarios studied in this section the CP-odd rate
asymmetry, equation~(\ref{eq:asymmetry}), is always quite small,
$|A_\mathrm{CP}| \lesssim 0.3\%$.
This due to the fact that in order to get a large 
asymmetry the effects from the absorptive phases in the Higgs propagators, 
$S_{H_i}$ as well as the phases in the couplings both have to be large. 
On the contrary, in the scenarios considered here the
phases in $S_{H_i}$ are always quite small and at the same time the mixing
between the neutral Higgs bosons is also typically small resulting in small
asymmetries. 
In order to get a non-negligible asymmetry one needs large phases 
from at least one of the propagators and at the same time large mixing 
between the neutral Higgs bosons, which seems difficult to achieve in 
the MSSM. 
In more general 2HDMs, large asymmetries should be possible in scenarios with
two resonant neutral Higgs bosons in the $s$-channel and large
mixing, although this remains to be verified.

\subsection{Resonant scenarios}

In SUSY parameter regions with $|\mu|, |A_t|,$ or $|A_b| > 4 M_\mathrm{SUSY}$ 
the dominant terms of the 1-loop corrections to the quartic couplings
in the Higgs sector \cite{Pilaftsis:1999qt}
can induce a large mass splitting
between the charged and neutral Higgs bosons
\cite{Akeroyd:2001in,mohngollubassamagan:2005}.
For example, in the scenario given in  table~\ref{resonantpar},
the CP-odd Higgs is 80 GeV heavier than the charged Higgs allowing
resonant production in the $s$-channel.

\begin{table}
\centering
\begin{tabular}{lcccccccccccc} \hline \hline
 &\multicolumn{11}{c}{MSSM parameters. All masses in GeV.}\\ 
  &
$m_{H^\pm}$ &
$\tan\beta$ &
$\mu$ & 
$M_L^3$ &
$M_E^3$ &
$M_Q^3$ &
$M_U^3$ &
$M_D^3$ &
$A_t$ & 
$A_b$ & 
$M_2$ &  
$m_{\tilde g}$  \\ \hline
Resonant scen. &
175 
& 11 
& 3300 
 & 500 
 & 500 
 & 250 
 & 250 
 & 400 
& 0 
& 0 
& 500 
 & 500 \\ 
Scan $\min$ &
100 &
1 &
1800 &
500 &
500 &
150 &
150 &
150 &
0 &
0 &
500 &
500 \\
Scan $\max$ &
450 &
40 &
3300 &
500 &
500 &
650 &
650 &
650 &
0 &
0 &
500 &
500 \\
Scan stepsize &
25 &
1 &
250  &
-- &
-- &
50  &
50  &
50  &
-- &
-- &
-- &
-- \\
\hline \hline
\end{tabular}
\caption{MSSM parameters for the resonant scenario as well as the 
range for the scan of parameters together with the stepsize.}\label{resonantpar}
\end{table}

In order to examine the perturbative stability of this scenario with
very large 1-loop corrections to the mass of the CP-odd Higgs boson we list
in table~\ref{massdiff} the masses of the Higgs bosons calculated with 
{\sc FeynHiggs}\footnote{We have added the partial decay width $\Gamma(H_i\to
H^\pm W^\mp)$ to the calculation of $\Gamma_{H_i}$ in {\sc FeynHiggs}.} 
at tree-level, at 1-loop order, and with all available corrections. 
For comparison, the Higgs masses in the maximal mixing
scenario are also given.
From the table it is clear that the higher order corrections have much smaller
impact than the leading ones suggesting that the perturbative expansion is
under control. We have also verified that similarly large mass splittings 
are obtained with {\sc CPsuperH} using all available corrections although 
not in precisely the same parameter points. The latter feature is due to 
this kind of resonant scenarios being
rather fine-tuned and therefore sensitive to differences in the
implementation and approximations used in the two programs.

\begin{table}[htbp]
\centering
\begin{tabular}{cc|ccc} \hline \hline
 & $m_h^\mathrm{max}$ & \multicolumn{3}{c}{Resonant scenario} \\
 & full & tree-level & 1-loop & full  \\ \hline
$m_{h^0}$ & 136 GeV & 89 GeV & 95 GeV  & 118 GeV  \\
$m_{H^0}$ & 151 GeV & 157 GeV & 188 GeV  & 168 GeV  \\
$m_{A^0}$ & \bfseries 151 GeV & \bfseries 155 GeV & \bfseries 246 GeV 
 & \bfseries 258 GeV \\
$m_{H^\pm}$ & \bfseries 175 GeV & \bfseries 175 GeV & \bfseries 175 GeV 
 & \bfseries 175 GeV  \\ \hline \hline
\end{tabular}
\caption{Masses of the Higgs bosons in the maximal mixing and resonant
  scenarios, calculated with {\sc FeynHiggs} at tree-level,
 at 1-loop order and using all available corrections. 
 }\label{massdiff} 
\end{table}

\begin{figure}
\centering
\includegraphics[width=7.5cm]{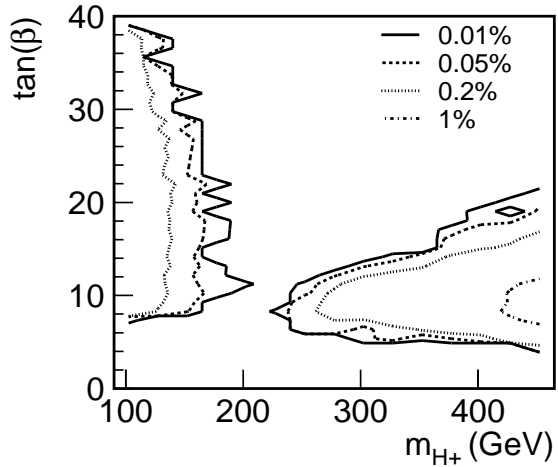}
\caption{Contour plot of the relative number of resonant scenarios to physical 
scenarios in each point. The lines are 0.01\%, 0.05\%, 0.2\% and 1\%. 
The region around $\tan\beta \approx 8$ contains more physical scenarios than the upper region leading to a total of 0.2\% resonant scenarios.
}\label{fig:resonantscan} 
\end{figure}

To get an indication of how fine-tuned this kind of resonant scenario is,
we have performed a scan over the relevant parameters as shown in 
table~\ref{resonantpar}.
Defining a resonant scenario via the relation $m_A > m_{H^\pm} + m_W$,
there is in fact a large range in both $\tan\beta$ and $m_{H^\pm}$ 
were such scenarios are found as can be seen in figure~\ref{fig:resonantscan}.
In each $\tan\beta$ and $m_{H^\pm}$ point about 9300 different scenarios are
tested. On average about one half of these correspond to physical scenarios 
and in turn about $0.2\%$ of the latter give resonant conditions, which
illustrates the level of fine-tuning in these scenarios. 
The dependence on $\tan\beta$ and $m_{H^\pm}$ is illustrated in 
figure~\ref{fig:resonantscan}.
At the same time, these scenarios
typically have relatively low squark masses so for large
$m_{H^\pm}$ ($\gtrsim 200$ GeV)
the decay to squarks becomes dominant and thus the specific 
analysis we present here is not suitable. At the same time, in scenarios with
light squarks there may also be a resonant enhancement in the gluon initiated
channel from squark loops as already discussed above. This occurs if the sum 
of the squark masses are close to threshold. In addition is is also possible in
this case to have enhancement from the $s$-channel resonance. Thus in the rare
situation that both these effects occur simultaneously, without opening the decay
channel of the charged Higgs into squarks, the signal could in fact be enhanced.
Finally, we note that the study of resonant scenarios  
also illustrates what could happen in a
general 2HDM, where the masses are more or less independent parameters.

\begin{figure}
\centering
\includegraphics[width=7.5cm]{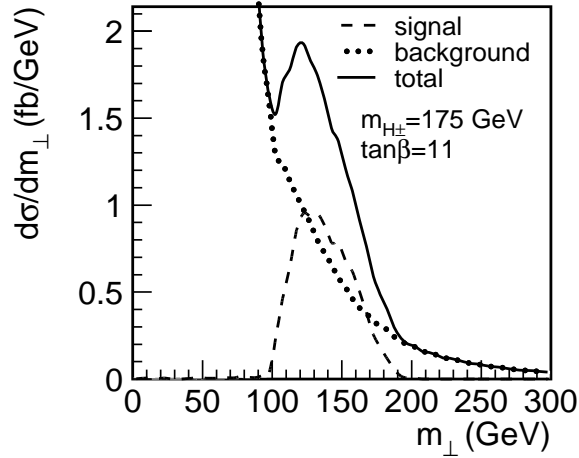}
\caption{Comparison of $m_\perp$ distribution between signal (for the resonant
  scenario with $m_{H^\pm} = 175$ GeV  and $\tan\beta=11$ 
  given in table~\ref{resonantpar}) and background
  with all cuts of table~\ref{tbcuts} except 
  $p_{\perp hj} > 50$~GeV and $p_{\perp sj} > 25$~GeV. 
  The dashed (dotted) curve denotes signal (background) and the
  solid curve is signal and background combined.}\label{resonant} 
\end{figure}

In case of resonant production, the $H^\pm$ and $W$ bosons are produced with
typically small transverse momenta. Thus it is favourable to loosen the cuts 
on the light jets from the $W$. Applying the basic and additional cuts from 
table~\ref{tbcuts}, 
except the cuts $p_{\perp hj} > 50$~GeV and $p_{\perp sj} > 25$~GeV 
on the light jets, we get an integrated cross-section of 52~fb 
for a charged Higgs boson mass of 175
GeV in the resonant scenario given in table~\ref{resonantpar}. 
Figure~\ref{resonant} shows the resulting $m_\perp$-distribution
 compared to the background. 
With an integrated luminosity of 300 fb$^{-1}$ and a $\tau$ detection
efficiency of 30\% we get a significance $S/\sqrt{B}=56$. 
For comparison, if we apply the harder cuts
$p_{\perp\tau_\mathrm{jet}} > 100$~GeV and 
 $\mbox{$\!\not \!p_{\perp}$}>100$~GeV
in this resonant case the significance is reduced drastically 
to $S/\sqrt{B}=0.2$ due to the typically 
small transverse momentum of the $H^\pm$-boson. Thus, in the case of harder 
cuts the resonantly enhanced cross-section is only of use if $m_{H^\pm}$
is large enough such that $m_{H^\pm}/2$ is well above $100$ GeV.

\begin{figure}
\centering
\includegraphics[width=7.5cm]{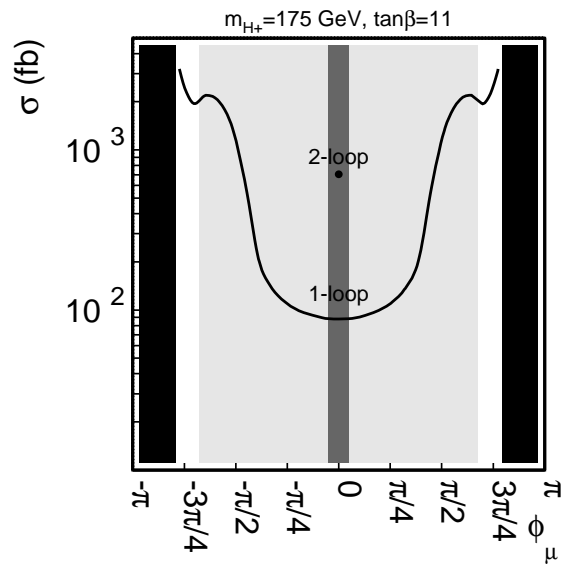}
\caption{Total cross-section, with the Higgs masses etc. calculated to 
one-loop, as a function of $\phi_\mu$ for
  the resonant scenario. For comparison the result with all available
  corrections for $\phi_\mu=0$ is also shown (labeled 2-loop). 
  The light shaded areas are in
  agreement with the constraints from $a_\mu$ and $\delta\rho_0$ and 
  the dark shaded ones 
  are also in agreement with the constraints from EDMs.
  In the black areas numerical instabilities occurred in the calculation of
  the Higgs masses and mixing matrix.}\label{fig:phases_res} 
\end{figure}

Finally we have also tried to investigate the phase dependence of the
cross-section in this resonant scenario.
Figure~\ref{fig:phases_res} shows the dependence of the cross-section on
$\phi_\mu$, where the Higgs masses, couplings and widths have been
calculated with {\sc FeynHiggs}  at one-loop accuracy\footnote{
A calculation with all available corrections is not possible here because the
phases lead to numerical instabilities.}.
Note, that the phase $\phi_{A_t}$ is irrelevant in this scenario
because $|A_t|=0$.
For comparison, the figure also 
shows the result with all available corrections in the case $\phi_\mu=0$. 
The very large phase dependence 
is due to the fact that the production goes
from non-resonant to resonant when varying the phase.
More specifically, as can be seen in table~\ref{massdiff}
we get $m_A = m_{H_3} = 246$ GeV for $\phi_\mu = 0$ in the 1-loop case, which
is below the resonant
threshold, whereas $m_{H_3}=342$ GeV for the largest values of $\phi_\mu$
where we got a stable result, which is clearly in the
resonant regime. Even in these resonant cases, the width of the $H_3$ is
typically small, $\Gamma/m \lesssim 0.02$, and consequently the CP asymmetry is
also small. However, we have not been able to make
a more thorough study of the sensitivity to these phases
due to numerical instabilities nor have we searched for resonant 
scenarios with large $\Gamma/m$ as well as large mixings
since this requires a dedicated study. 

\section{Summary and conclusions}

In this paper we have studied the viability of detecting charged Higgs bosons
produced in association with $W$ bosons at the upcoming LHC experiments. 
Since our study is of exploratory character we have stayed on the parton level,
although with appropriate smearing of momenta, and leave the inclusion of 
parton showering, hadronisation, full simulation of the detector, etc.
for future studies. On the same vain we have not tried to include any higher
order corrections in the production cross-section. The only exception is 
that we use running quark masses in the Yukawa couplings, which gives 
a cross-section that is almost a factor three smaller compared to when
using the pole-mass, in agreement with NLO QCD~\cite{Hollik:2001hy}. 
Other higher order corrections on the production cross-section, such as the
choice of factorisation scale, are expected to be $\mathcal{O}(10-20\%)$.

As we have shown in this paper, using the leptonic
decay of the charged Higgs boson and hadronic $W$ decay, giving rise to 
the signature
$\tau_\mathrm{jet} \mbox{$\not \!p_\perp$} + 2 \textrm{ jets}$,
it is possible
to design appropriate cuts against the irreducible Standard Model 
background from  $W$+ 2 jets production. 
This is in contrast to earlier studies using hadronic
decays of the charged Higgs bosons, which was found to suffer from too large
irreducible background due to $t \bar t$ production~\cite{Moretti:1998xq}.
At the same time we find that the significance of 
the resulting signal does depend on the 
assumptions made on the cuts needed against reducible backgrounds
(mainly from QCD and detector misidentifications). 

In the standard maximal mixing scenario
of the MSSM, using the softer
cuts ($p_{\perp\tau_\mathrm{jet}} > 50$~GeV and $\mbox{$\!\not \!p_{\perp}$} > 50$~GeV)
we find a viable signal in the case of large $\tan\beta$
($\gtrsim 30$) and intermediate charged Higgs masses 
($150~\textrm{GeV} \lesssim m_{H^\pm} \lesssim 300$~GeV), whereas 
with the harder cuts ($p_{\perp\tau_\mathrm{jet}} > 100$~GeV and 
$\mbox{$\!\not \!p_{\perp}$} > 100$~GeV) 
we only find a viable signal for 
$\tan\beta \gtrsim 50$ if in addition an appropriate upper cut on 
$m_\perp=\sqrt{2p_{\perp\tau_\mathrm{jet}} \mbox{$\not \!p_\perp$} 
[1-\cos(\Delta\phi)]}$ is applied. Thus, in the best case 
the associated charged Higgs and $W$ boson production could serve as a
complement to production in association with top quarks in the difficult
transition region $m_{H^\pm} \sim m_t$ although only for large $\tan\beta$,
but before one can draw any firm 
conclusions the effects of reducible backgrounds
have to be studied in more detail.

In MSSM scenarios with less or no mixing in the third generation squark sector
we get similar results as in the standard maximal mixing scenario.
In fact, the 
differences compared to the maximal mixing scenario are in both cases
smaller than the $\mathcal{O}(10-20\%)$ effects 
we expect from  higher order electroweak and QCD corrections 
as well as other variations of the SUSY scenario. 
Similarly we have also found that 
the cross-section depends only weakly on the CP-violating phases
of the SUSY parameters, even those of $A_t$ and $\mu$, which result
in a mixing between the CP-even and CP-odd Higgs bosons.
This is a general feature that will be true as long as
the differences in the neutral Higgs bosons masses are small compared to the 
charged Higgs plus W-boson mass (more
specifically $m_{H_i}^2 - m_{H_j}^2 \ll (m_{H^\pm}+m_W)^2 - m_{H_i}^2$) even if
the Higgs mixing matrix is highly non-diagonal.
The phases of $A_t$ and $\mu$ also lead to differences in the masses of the
neutral Higgs bosons. In fact, we have seen that the main effect on the 
cross-section comes from these kinematic effects of the Higgs masses
and not from the changes in the couplings due to the phases.

We have also studied a class of special resonant scenarios  
where $m_{H_i} \gtrsim m_{H^\pm}+m_W$ for one of the neutral 
Higgs bosons (the CP-odd Higgs $A$ in the MSSM with real parameters)
leading to resonant production in the $s$-channel.
This results in a very large enhancement of the total cross-section (up to a 
factor 100) in the region of intermediate $\tan\beta$ ($\sim 10$). 
However, due to the different kinematics of resonant production, the
significance of the signal depends very strongly on the 
$p_{\perp\tau_\mathrm{jet}}$ and $\mbox{$\!\not \!p_{\perp}$}$ cuts. 
With the softer cuts we get a significance of order 50 in 
the case of  $m_{H^\pm} = 175 $ GeV, whereas with the harder
ones it drops to $0.2$. Another problem
with these type of MSSM scenarios is that for larger charged Higgs masses
($m_{H^\pm} \gtrsim 200 $ GeV) the decays to squarks opens up which would 
require a different type of analysis.

One may also worry about
the perturbative stability of these resonant scenarios with very large
one-loop corrections
to the mass of the CP-odd Higgs boson.
 However, comparing the one-loop and available
two-loop corrections,
as was done in Table \ref{massdiff},
this does not seem to be the case.
When making a sparse scan of parameters
we find similar
scenarios in a large range of $\tan\beta$ and $m_{H^\pm}$
and we also get similar results when using {\sc CPsuperH}
instead of {\sc FeynHiggs} although not precisely at the same points in
parameter space.
We have also found that in these resonant scenarios the cross-section can 
have a large
dependence on the CP violating phases of the SUSY parameters but we have not
found any appreciable CP asymmetries. However, we have not been able to make
a more thorough study of the sensitivity to these phases which would require a
separate study.

Finally, it should be emphasized that our study is specifically for different
MSSM scenarios and that the conclusions may change in other models. For example, 
in a general 2HDM, the resonant enhancement is more natural since the Higgs
masses are independent of each other. In addition there will be no charged
Higgs decays to squarks as we found in the MSSM thus making it possible to 
have a clear signal for large charged Higgs masses even with the harder
cuts against reducible backgrounds. Similarly nonminimal supersymmetric models
with larger Higgs sectors may also
offer more natural possibilities for resonant enhancement. 

\section*{Acknowledgements}

We would like to thank Gunnar Ingelman for comments on the manuscript.
This work has been supported by the G\"oran Gustafsson Foundation.
S.H.\ is supported by the German Federal Ministry of Education and Research
(BMBF) under contract number 05HT4WWA/2. J.R. is supported by the Swedish
Research Council, contract number 629-2001-5873.

\bibliography{chargedhiggs}

\end{document}